\documentclass[a4paper,9pt]{extarticle}

\usepackage[UKenglish]{babel}

\usepackage{newfloat}
\usepackage{float}
\usepackage{soul}
\usepackage[table, svgnames]{xcolor}
\usepackage[bookmarksdepth=3]{hyperref}
\usepackage{bookmark}
\usepackage{etoolbox}
\usepackage[utf8]{inputenc}
\usepackage[style = english]{csquotes}
\usepackage{pdflscape}
\usepackage{afterpage}
\usepackage{pdfpages}
\usepackage{datetime2}
\usepackage{fancyhdr}
\usepackage{multicol}
\usepackage{multienum}
\usepackage{enumitem}

\usepackage{alertmessage}

\newbool{twoSide}
\makeatletter
\if@twoside%
	\booltrue{twoSide}
\else%
	\boolfalse{twoSide}
\fi%
\makeatother

\newbool{hasChapters}
\makeatletter
\@ifclassloaded{report}{%
	\booltrue{hasChapters}
}{%
	\boolfalse{hasChapters}
}
\makeatother

\ifbool{twoSide}{
	\hypersetup{hidelinks, colorlinks, urlcolor=Black, linkcolor=Black, citecolor=Black}
}{
	\hypersetup{hidelinks, colorlinks, urlcolor=DodgerBlue, linkcolor=MidnightBlue, citecolor=DarkOliveGreen}
}

\newcommand{\AutorefChap}[1]{\hyperref[#1]{Chapter~\ref*{#1}}}
\newcommand{\AutorefAlg}[1]{\hyperref[#1]{Algorithm~\ref*{#1}}}
\newcommand{\AutorefApp}[1]{\hyperref[#1]{Appendix~\ref*{#1}}}
\newcommand{\AutorefSec}[1]{\hyperref[#1]{Section~\ref*{#1}}}
\newcommand{\AutorefSubsec}[1]{\hyperref[#1]{Subsection~\ref*{#1}}}
\newcommand{\refEqP}[1]{\hyperref[#1]{(Eq.~\ref*{#1})}}
\newcommand{\refEq}[1]{\hyperref[#1]{Eq.~\ref*{#1}}}

\ifbool{twoSide}{
	\newcommand{\leftMargin}{3.08cm} 
	\newcommand{\rightMargin}{2cm}
}{
	\newcommand{\leftMargin}{2.54cm} 
	\newcommand{\rightMargin}{2.54cm}
}  
\newcommand{\topmarginCust}{2.54cm}
\usepackage[top = \topmarginCust, bottom = 2.54cm, left = \leftMargin, right = \rightMargin]{geometry}

\let\oldhl\hl
\renewcommand{\hl}[1]{\noindent\parbox{\linewidth}{\oldhl{#1}}}

\usepackage{enumitem}
\usepackage{lmodern}
\usepackage[T1]{fontenc}
\usepackage{palatino}

\newlength{\mytextsize}

\usepackage[backend=biber, style=apa, dashed=false, doi=false, isbn=false, natbib=true, giveninits=true, uniquename=init, sorting=nyt, autocite=plain, hyperref=true, backref=true]{biblatex}
\DeclareNameAlias{author}{sortname}
\DeclareNameAlias{editor}{sortname}
\DeclareNameAlias{sortname}{last-first}
\DeclareNameAlias{default}{last-first}
\renewcommand{\cite}[1]{\citep{#1}}

\usepackage{amsmath}
\usepackage{bm}
\usepackage{array}
\usepackage{amssymb}
\usepackage{amsfonts}
\usepackage[linesnumbered,ruled,noend]{algorithm2e}
\SetAlgoCaptionLayout{small}
\SetAlFnt{\small}

\newcommand{\vect}[1]{\mathbf{#1}}

\newcommand{\vecut}[3]{\vect{#1}_{#2}(#3)} 
\newcommand{\vecu}[2]{\vect{#1}_{#2}}

\newcolumntype{R}{>{$}r<{$}}
\newcolumntype{L}{>{$}l<{$}}

\usepackage{graphicx}
\usepackage{epstopdf}
\usepackage{subcaption}
\usepackage{tabu}
\usepackage{booktabs}
\usepackage{longtable}
\usepackage{hhline}
\usepackage[font=small, format=hang]{caption}
\usepackage{capt-of}
\usepackage{multirow}
\usepackage{wrapfig}
\usepackage{makecell}

\newcommand{\mc}[3]{\multicolumn{#1}{#2}{#3}}

\definecolor{lightGray}{RGB}{230,230,230}

\usepackage{cleveref}

\usepackage{xargs}
\newcommandx*\cvGen[4][2=,3=,4=,usedefault]{#1_{u#4}^{i#2,j#3}}
\newcommandx*\cvRight[3][2=,3=,usedefault]{#1_{u,\text{right}}^{i#2,j#3}}
\newcommandx*\cvLeft[3][2=,3=,usedefault]{#1_{u,\text{left}}^{i#2,j#3}}
\newcommandx*\cvTop[3][2=,3=,usedefault]{#1_{u,\text{top}}^{i#2,j#3}}
\newcommandx*\cvBottom[3][2=,3=,usedefault]{#1_{u,\text{bottom}}^{i#2,j#3}}

\newcommand{\acp}{\alpha_\text{cp}}
\newcommand{\bcp}{\beta_\text{cp}}
\newcommand{\add}{\alpha_\text{dd}}

\usepackage[pagewise, mathlines, displaymath]{lineno}
\usepackage{authblk}
\usepackage[frozencache=true, newfloat=true,cachedir=minted-cache]{minted}
\newenvironment{code}{\captionsetup{type=listing}}{}
\SetupFloatingEnvironment{listing}{name=Source Code}

\newenvironment{opCountTabu}[1][0.45]
{\begin{tabular}{rRlRp{#1\linewidth}}
	}
	{
\end{tabular}}

\newenvironment{memopCountTabu}[1][0.42]
{\small\begin{tabular}{p{0.2\linewidth}lRRRp{#1\linewidth}}
	\textbf{Action} & \textbf{Type} & \textbf{Count} & \textbf{Mp.} & \textbf{Costs} & \textbf{Details}\\
	\hline
	
}
	{
\end{tabular}}

\newenvironment{opCountTabuMul}[1][0.32]
{\small\begin{tabular}{rlRRRRp{#1\linewidth}}
	\textbf{Line} & \textbf{Operation} & \textbf{Count} & \textbf{Op. cost} & \textbf{Mp.}  & \textbf{Costs} & \textbf{Details}\\
	\hline
	}
	{
\end{tabular}}

\modulolinenumbers[1]

\graphicspath{{./Graphics/}}
\bibliography{How_fast_is_my_model_APS_preprint_Sparnaaij2025}

\title{How fast is my model? \framework{}: A framework to systematically assess the computational speed of pedestrian models}
\author[1,*]{Martijn Sparnaaij}
\author[1]{Dorine C. Duives}
\author[1]{Serge P. Hoogendoorn}
\affil[1]{\small Department of Transport \& Planning, Delft University of Technology, Stevinweg 1, 2628 CN Delft, The Netherlands}
\affil[*]{\small Corresponding author: Martijn Sparnaaij, M.Sparnaaij@tudelft.nl}

\date{\small Version: \today}

\newcommand{\framework}{APS}
\newcommand{\frameworkLong}{Assess Pedestrian model Speed}

\begin{document}
 
	\vspace{-5cm}
	\maketitle
	
	
	\begin{abstract}
		A pedestrian model’s computation speed impacts the model applicability. However, little attention has been given to this model property in the field of pedestrian dynamics modelling. As such, no framework exists to guide the systematic analysis of a pedestrian models' computational speed. This contribution presents the \framework{} framework (\frameworkLong{} framework), a framework to determine the speed of (pedestrian) dynamics models. \framework{} features three methods to assess the computational speed of an algorithm, each tailored specifically to the use case of pedestrian models. It also provides guidance in choosing the proper method or methods depending on the goal and requirements of the analysis. \framework{} also includes a new procedure to produce test cases. By using multiple test cases, the framework ensures that the dependency of a model’s computational speed on the simulated scenario is assessed systematically.
	\end{abstract}
	
	\begin{center}
		\vspace{5mm}
	\end{center}

	\section{Introduction}
	\label{sec:introduction}
	
	State-of-the-art pedestrian models have many potential applications such as evacuation studies, supporting the design of pedestrian facilities or modelling the spread of an airborne virus like Covid-19. Many different models and model types have been developed over the past decades. Reviews such as \cite{Papadimitriou2009,Duives2013,Martinez-Gil2017} show that these models vary widely on multiple facets. They describe the walking behaviour at different levels (macroscopic/mesoscopic/microscopic) and they apply different modelling paradigms, such as Cellular Automata \cite{Blue1999}, Social Force \cite{Helbing1995}, Discrete Choice \cite{Antonini2006} and Cell Transmission \cite{Hanseler2014}.
	
	With so many different applications and so many different models, the question often is: ‘Which models are best suited for which application?’ This question falls apart into parts, namely in 1)'is a model producing realistic result for the considered simulation scenario(s)?' and 2) 'is a model fast enough for my current purpose?'. The first question relates to the model’s validity for a given set of scenarios. That is, how accurately can the model reproduce the relevant crowd behaviours and phenomena for a given application? And, is this level of accuracy sufficient for the given application? This question and the related questions regarding verification and calibration have gotten ample attention from the pedestrian modelling community. Many papers presenting pedestrian simulation models present some form of model assessment that provides insights into a model’s validity. In addition, multiple frameworks have been developed in the last decade that give guidelines to systematically test a model’s validity, amongst others \cite{IMO2007,Ronchi2013,RiMEA2022}.
	
	The second part of the question relates to the computational effort that a model requires to produce its results. That is, 'can the model produce the results within the required time given the available computational resources?’ Or simply said, is the model fast enough? For some applications, such as real-time forecasting and decision support or (real-time) optimisation, low computational effort is critical. In other cases, for example, a faster model can improve a scientist's or simulation engineer's ability to perform uncertainty quantification. For instance, by allowing an engineer to run more simulations within the same amount of time using a given amount of computational resources than a slower model would allow. Contrary to validity, the question of the computational speed of pedestrian models has received very little attention. Papers rarely provide information about a model’s computational speed or effort, with some notable exceptions \citep[e.g.][]{ Tordeux2018,Wagoum2012,Sung2004,Chooramun2019}, and to the authors’ knowledge no systematic way of assessing the computational speed has been published.
	
	The lack of a systematic way to assess a pedestrian model’s speed hinders model development and model application. For instance, it becomes more difficult to answer the question of which pedestrian model is best suited for which application. To fill this gap, we present \framework{} which provides a systematic way of assessing the computational speed of pedestrian models. We will show that systematic means that a clearly defined and described method is used that considers the variety of variables that impact the speed of pedestrian models and that considers the limitations of this method. 	
	
	Before we present \framework{}, we first define what we mean by a pedestrian model. Fully functional pedestrian models are comprised of various modules featuring sets of complex algorithms. Each of the modules is responsible for different levels of pedestrian behaviour (i.e., the strategic, tactical, and operational \cite{Hoogendoorn2003}) or basic functionalities to initialize and run the simulation and save the simulation results. Generally, a pedestrian model will contain at least six modules performing the following functionalities or describing the following behaviours:
	\begin{itemize}
		\item The walking behaviour (operational behaviour). This is the combination of the path-following behaviour and the collision avoidance behaviour.
		\item The route choice behaviour (tactical behaviour). This behaviour is either modelled or given as input in the form of a static floor field for example.
		\item The activity choice (strategic behaviour) and scheduling behaviour (tactical behaviour). Again, this is not necessarily modelled but can also be given as input.
		\item The model boundary processes. (source and sink). This is the algorithm that manages where and when pedestrians enter the simulation based on the input and handles pedestrians leaving the simulation.
		\item The model initialization from the input
		\item The saving of the data to a file or memory
	\end{itemize}
	
	Each of these parts is, up to a certain degree, interchangeable (e.g. most route choice models can be made to work with most walking models) and/or is independent of the other parts (i.e. what data is saved and how does not follow directly from the design of a walking behaviour model or a route choice model). The walking behaviour module is the part of pedestrian models that receives most attention from the pedestrian modelling community. Furthermore, most pedestrian models presented in the literature only focus on the walking behaviour part and to a limited degree describe the other parts if this is done at all. That is, they generally don't describe the route choice model, the source and sink behaviour etc. So, given that the walking behaviour part of the models has received the most attention and consequently has the largest variety of models that are well described, we solely focus on the walking behaviour part of pedestrian models in this paper. In the remainder of the paper, we simply refer to the walking behaviour model of a comprehensive pedestrian simulation model as a ‘pedestrian model’.
	
	The remainder of this paper is built up as follows. In the next section (\autoref{sec:framework}) we introduce the framework for the systematic determination of the computational speed and provide a definition for the computational speed of a pedestrian model. Next, in \autoref{sec:methods} we introduce the speed methods we have selected to be part of the framework. Following that, we introduce the method to select and define the test cases in \autoref{sec:test_cases}. Then, in \autoref{sec:showcase} we apply this new framework to a social force model to showcase it. Lastly, in \autoref{sec:conclusions} we present our conclusions and discuss the limitations and opportunities for future research.	
		
	\section{\framework{}: the framework to systematically assess the computational speed of pedestrian models}
\label{sec:framework}

In this section we present the \framework{} framework. However, before we present our new framework, it is important to define what we mean by the computational speed of a pedestrian model. So, first we define computational speed followed by the introduction of \framework{}. 


\subsection{Defining computational speed}
\label{subsec:framework:complexity_definition}
The speed of an algorithm is formally defined as the \emph{time computational complexity} of that algorithm. Computational complexity is defined as the amount of resources required to run an algorithm for a given input size \cite{Sedgewick1996}. Commonly, the resources are the computation time and the memory and in our case, the algorithm is a pedestrian model. Within this paper, we focus solely on the resource time and ignore the memory. This is because: 1) The amount of memory is generally not a limiting factor for most pedestrian models; even for very large scenarios given the amount of memory in current-day computers. And 2), the focus of this paper is on assessing the speed of pedestrian models, and thus on the time computational complexity.

Within this paper, we adopt the common definition of time computational complexity stated above and extent it to fit better to the context of (pedestrian) simulation models. We define the time computational complexity of a pedestrian model as:  \emph{the computational effort per unit of simulated time for a given input size}.

The computational effort reflects the amount of (time) resources required to perform a simulation step of the model. The amount of resources is reflected by a different quantity for different methods. As \autoref{sec:methods} will show, our new framework contains three different methods, two theoretical and one empirical, for determining the computational effort of a model. For the two theoretical methods, the amount of resources is quantified by the number of operations performed per simulated unit of time. For the empirical method, the amount of resources is quantified by the computation time per unit of simulated time. \AutorefSec{sec:methods} will explain in more detail why we make this distinction.

Using the unit of simulated time is important as this enables making comparisons between different pedestrian models that use different time steps or between time-step-based and event-based models. We use 1 second as the unit of simulated time within this paper as most pedestrian models have a time step in the order of (a fraction) of seconds. In the remainder of the paper, we simply refer to the time computational complexity as the \emph{'computational speed'} or simply the \emph{'speed'} of a model.

In addition, it is important to define the concepts "input" and "input size". The input to a comprehensive pedestrian model is a scenario defining contextual characteristics and elements, such as the infrastructure and the demand pattern. However, this is not the input to the algorithm that manages the operational walking dynamics in a model. The input to this part of a pedestrian model is (a part of) the current state of the model. This state could feature, among other things, the collection of all active pedestrians, the collection of all obstacles, and/or the collection of all cells depending on the exact model. The input of a model can consist of multiple elements. For example, the input to a simulation step of the social force model consists of the collection of all active pedestrians and the collection of all obstacles.

The size of the input is the size of each input element. Because the input can consist of multiple elements, the input size is not necessarily singular. If we take the example of the social force model again, the input to a simulation step of the model has two elements (pedestrians and obstacles) and thus two input size (the number of pedestrians and the number of obstacles). Each of these input sizes can impact the speed of the model in a different way and to a different degree. Therefore, we need to analyse the impact of each input element and combinations of input elements and their input sizes separately.

\subsection{The \frameworkLong{} framework }
\label{subsec:framework:overview}
The main goal of our new framework is to provide a guideline to systematically analyse the speed of pedestrian models. To develop the framework, we combine our observations about the common ways in which these analyses have been performed in the field of pedestrian dynamics so far, observations about the inputs to pedestrian models and observations about the different methods one can use to obtain the speed of an algorithm.  

Firstly, we observe that the input to pedestrian models can consist of multiple elements whose size impact the model's speed (see \autoref{subsec:framework:complexity_definition}). We also observe that different (types of) pedestrian models have different input elements. For example, a social force model has completely different input elements compared to a cell-based macroscopic model. The list of pedestrians and the list of obstacles are input elements of the social force model but not of a cell-based macroscopic model. These cell-based macroscopic model do have the list of cells that cover the walkable space as their input which is, in turn, not an input to microscopic models. So, the speed of microscopic models depends directly on the number of pedestrians and obstacles in a simulation but not directly on the size of the walkable space whilst for cell-base macroscopic models this is vice versa.

From these two observations, we conclude that our systematic framework needs to able to assess the impact of multiple input elements on a model's speed. We tackle this problem by introducing a procedure that systematically creates of a set of pre-described test cases. We define a test case as a description of a specific state and space of pedestrian models (i.e. a specific set of inputs, their accompanying sizes and input values to pedestrian models). So, each test case has a specific input size for each input element. Then multiple test cases are created with varying input sizes for the various elements forming a set of test cases. The core principle is that this set of test cases includes all the input elements that impact a model's speed. This ensures that a complete insight is obtained into how and to which degree each elements of the input impact the model's performance.

\AutorefSec{sec:test_cases} will present the procedure to create the set of test cases for a model in detail. Furthermore, it also will present guidance on how to create the set of test cases when comparing multiple models with different input elements.

Our third observation is that there are different reasons why someone wants to gain insight into the speed of pedestrian models. One goal could be to compare different models on a fundamental level. That is, answering the question of which model is faster in which scenario, independent of how optimally the models are implemented compared to each other, what hardware is used or how easily they can make use of techniques such as parallelization. Another goal could be to analyse the speed of a single model to obtain insight into what scenario elements mainly determine the model’s speed, and thus for what scenarios this model is well suited (in combination with information about its validity) and for which scenarios this model might not be the optimal choice. A third goal could be to compare or analyse implemented model instances in order to gain insight into their computation time or effort for a given hardware setup featuring different scenarios. 

No single method can accomplish all these different goals (see \autoref{sec:methods} for details). Therefore, our comprehensive framework should contain multiple methods to cover these three different goals. Furthermore, our framework should provide guidance regarding how to choose the right method or methods to asses a model's speed depending on the goal or goals of the analysis. And lastly, it should clearly describe the limitations of each of these methods to ensure that the result are interpreted correctly.

Based on these observations we designed a framework consisting of the following steps:
\begin{description}
	\item[Step 1:] Based on your goals and the requirements regarding the type of your analyses, select one or more of the three methods to determine the computational speed (see \autoref{sec:methods}). 
	\item[Step 2:] For each model, provide the model specification (see \autoref{sec:methods}).
	\item[Step 3:] For each model and method, determine the set of relevant test cases (see \autoref{sec:test_cases}).	
	\item[Step 4:] Apply each method to each model using the combination of test cases to derive the speed of each model.
\end{description}

\begin{figure}[htb]
	\centering
	\includegraphics[width=\linewidth]{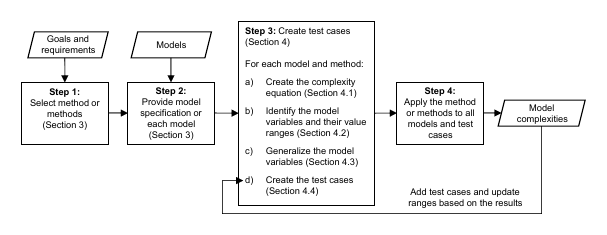}
	\vspace{1mm}
	\caption{Overview of the framework for obtaining the speed of pedestrian models}
	\label{fig:framework}
\end{figure}

\autoref{fig:framework} presents a visual overview of the framework. For more information on the implementation of the steps, we refer the reader to \AutorefSec{sec:methods} and \autoref{sec:test_cases}. \AutorefSec{sec:methods} presents all three methods we include in our framework and also how to apply them and what their strengths and limitations are. Here, \autoref{subsec:methods:choose} provides a summary of these strengths and limitations and provides further guidance on how to choose the right method or methods.
In section \autoref{sec:test_cases} we provide a detailed method that guides a user in compiling the right set of test cases for each model and method.

	\section{Methods to determine the computational speed of pedestrian models}
\label{sec:methods}

The main goal of any method to assess a model's speed is to take a model and a set of test cases and determine the model's speed in relation to the inputs and input sizes defined in the test cases. In the rare cases that the speed of a pedestrian simulation model is analysed in literature, one out of three methods is used. Firstly, the empirical method. This method generally involves running the model (or models) for a particular scenario whilst varying the number of simulated pedestrians \citep[e.g.][]{Paris2006,Xi2011,Banerjee2008}. The empirical method produces a relation between the model run time and the number of pedestrians that are simultaneously present in the simulation. Secondly, the asymptotic method. This method uses the concept of the Big-O notation to identify how the model’s speed scales with respect to a certain input type. Also in this case, often the relation between Big-O and a single input, such as the number of pedestrians, is evaluated \citep[e.g.][]{Wagoum2012,Seyfried2011,Tordeux2018}. Lastly, \cite{Tordeux2018} define their model's speed using a detailed equation. They derived this equation by determining how many operations each computation of the model required in relation to the number of pedestrians and combining these parts into a single equation.

Each of these three methods to determine the computational speed of a model has different properties, strengths and weaknesses. There is not one method that suits all goals and all purposes. As a result, the method one should adopt is dependent on ones particular goal(s), and the types of insight(s) you want to obtain about a model's speed (see \autoref{subsec:framework:overview}). Therefore, we include all three methods in our framework. In the remainder of the section, we introduce and define each method in detail and describe how it should be applied by use of an example. Furthermore, we also define for each method what the model specification is that needs to be provided.

%
%

\subsection{The empirical method}
\label{subsec:methods:empirical}
The empirical approach is commonly used to assess a pedestrian model’s speed. Generally, a run time or frame rate is reported in relation to the number of pedestrians; whereby the latter is the independent variable. However, the method used to obtain these results is typically not described in detail. For example, it is not described which modules of the comprehensive pedestrian model are included in the run time measurement. Does the process only feature the walking behaviour, or does this also include the route choice, and/or the model boundary processes? Besides that, it is often not described how the actual computational speed was determined. That is, how is the relation between run time and the input size (generally the \# pedestrians) determined? Is this done by running one scenario with N pedestrians and measuring the total running time of this scenario? Or, has a scenario with a varying number of pedestrians been run and is the run time of each individual time step measured and coupled to the number of pedestrians in the simulation at that time step?

In short, this lack of detailed description of the employed method makes it hard to interpret the results and compare the results of different model studies. To alleviate these issues, we provide a detailed description of the empirical method that we include in our framework. In our method, we use the following core principles:
\begin{enumerate}
	\item \emph{We only record the run time for the part of the algorithm that governs the walking behaviour.} This is in line with our definition of the speed of a pedestrian model and ensures that the definition of model speed is unambiguous with respect to what is being measured and what the results of this computation represent. 
	\item \emph{We initialize the model with a state defined by a test case and measure the run time of one single time step} (or an event-based equivalent). The advantage of running only one single time step is that we can precisely control the input to the algorithm and hence precisely measure the impact of different inputs and input sizes.
	\item \emph{We perform replications to limit the influence of disturbances on the run time measurement} caused by, for example, other processes running on the same computer (which we advise to limit to an absolute minimum during the simulations).
\end{enumerate} 

Based on these three core principles, we designed an algorithm that forms the core of our empirical method (see \autoref{alg:empirical_method}). The inputs to this algorithm are the list of models you want to analyse and the set of relevant test cases. First, we create the list of experiments, where an experiment is a combination of a test case and a model. Then, a list of N replications per experiment is created and we randomly order this list of all experiments and replications. After running all replications of all experiments, a check is performed to see if the mean run time of each experiment has converged to a stable mean run time. For each experiment whose run time has not yet converged, we perform another N replications. This second step is repeated for all non-converged lists until the run time of each experiment has converged. 

Please note, that we randomly order the list to decrease the impact of any disturbances caused by other processes that are running on the computing platform simultaneously. Additional ways to limit the variance between measurements are to limit the number of other processes running on the computer to an absolute minimum and to set the processor frequency to a fixed value. Furthermore, it is important to document how many CPU cores a model has used during the time step when comparing models (or if it has used the GPU). This is important as this could also explain any differences found between the run times of different models. Ideally, the simulations are all run using a single CPU core. This allows for the best comparison between different model instances. This is unless you are interested in how different uses of computer resources by different models or model implementations affect the run times. 

Lastly, the measured mean run times need to be converted to a mean run time per simulated second (in line with our definition of speed). This is simply done by multiplying the run time by $1/\Delta t$ where $\Delta t$ is the time step of the model in seconds. A Python implementation of this method can be found at \cite{Sparnaaij2025a}.

\begin{algorithm}[H]
	\SetAlgoLined
	Create the list of experiments from the combination of test cases and models\\
	\While{run time of all experiments has not converged}{
		Create a randomly ordered list with N replications per non-converged experiment\\	
		\ForEach{experiment in the list}{
			Initialize the model using the experiment's state\\
			Run a single time step of the model and measure the run time\\					
		}
		Compute convergence criteria per experiment
	}
	Convert the measured mean run times to mean run times/simulated second 
	\caption{Pseudocode describing the Empirical method.}
	\label{alg:empirical_method}
\end{algorithm}

The approach presented in \cite{Ronchi2013} is adopted to test if the mean run time has converged for an experiment. This approach states that the mean run time of an experiment has converged if:

\begin{equation}
	\frac{\mu_i}{\mu} < \varepsilon \: \forall \: i \: \in \{1,2,...,M\}, \quad \mu_i = \frac{\sum_{j=0}^{n-i}t_{run;j}}{n-1} 
\end{equation}

where $\mu$ is the mean run time using the run times from all $n$ replications, $M$ is the number of consecutive means that are checked, $\varepsilon$ the allowable relative error, $n$ is the total number of replications performed so far and $t_{run;j}$ is the run time of replication $j$. Or in other words, $M$ mean run times ($\mu_i$) are computed based on the run times of all replications except for the last $i$ replications. Each of these mean run times is divided by the mean of the run times of all replications ($\mu$) resulting in $M$ relative error values. If all $M$ relative error values are smaller than the set threshold value ($\varepsilon$), the mean run time has converged to a stable value. In our case, we adopted $M = 10$ and $\varepsilon = 1\%$, values similar to those chosen in \cite{Ronchi2013}.

The model specification you need to provide for this method is the operational program code of the model. Running this code for all test cases and replications results in a mean run time per experiment. That is, the result is a mean run time for a specific combination of inputs and accompanying input sizes. By means of these insights you can determine the relationship between the model run time and a specific input by combining the data of different experiments. For example, by combining experiments with different numbers of pedestrians but otherwise a state that is exactly the same, the relationship between the number of pedestrians and the model run time can be determined. A common way to analyse the results of the Empirical method is by plotting the run time using different inputs and their sizes as the independent variable. For example, plotting the relationship between the number of pedestrians for different numbers of obstacles (i.e. a line per number of obstacles) and the run time for different numbers of obstacles for different numbers of pedestrians. Examples of the output of the Empirical method can be found in \autoref{sec:showcase}.

The current definition of the method focuses on time-step-based models. Though these types of models represent the overwhelming majority of all pedestrian models, some models use an event-based approach. The empirical method, as we have defined it, can be applied to event-based models as well, by adopting the following slight adaptation to the method described above. To get a run time per simulated second, measure the time it takes to process all events that take place to simulate one second simulated time. In the case that no events happen within this one second, or so few event happen that the run time cannot be measured  because its too short, we advice another procedure. In this case, process the first $N$ events that take place, measure the run time it takes to process these $N$ events and divide this by the simulated time that passed between the start of the simulation and the last processed event (event number $N$). We cannot provide a number for $N$ as it depends on the model and the computing platform. Therefore, we advice to test a few values for $N$ to check what value produces measurable results.

Please note! An important property of the empirical method is that it is computing platform-dependent. With computing platform we mean the combination of hardware and software that is used to run the simulations. The hardware includes, among other things, the CPU, the motherboard, the clock speed of the memory. The software includes the operating system, the compiler and the programming language used to program and run the model. This platform dependency has several implications, both with respect to how one should use the method and how to interpret the results. The first implication relates to the degree to which model complexities can be compared when these have been obtained using different computing platforms. The absolute values of the run times are heavily dependent on the computing platform. Therefore, the comparison is in those cases restricted to the relative growth rates in relation to the inputs. To compare the complexities on a more detailed level, the results should be obtained using the same computing platform. 

Another implication of the platform dependency is that the results are very implementation-dependent. That is, different implementations of the same model can lead to different results because certain coding patterns or libraries are more efficient than others. The efficiency also depends on the exact programming language used. For example, in Matlab and Python loops can sometimes be vectorised, which increases the code's performance and decreases the run time. Therefore, if this optimization is applied to one model implementation but not the other, any differences in their empirical speed might be just due to the different levels of optimization of the implementation and not due to fundamental differences in the computational speed of the models. Similarly, some programming languages are generally faster than others. For example, compiled languages (e.g. C++) tend to be faster than interpreted languages (e.g. Python).

Therefore, any conclusions regarding the speed of (pedestrian) simulation models, obtained using any empirical method (including ours), are always bound to the particular implementation of the model. To obtain insight into any fundamental differences between model complexities, one should take care that all implementations are of a similar optimization level and written in the same programming language. From this, it also follows that a detailed record of the exact implementation should always be provided with the speed results to enable a proper interpretation and comparison of the results. Yet, the dependency on the implementation also has an advantage. Namely, it is easy to compare how different implementations of one particular model impact its computational speed.

\subsection{The asymptotic speed method}
\label{subsec:methods:asymptotic}

The asymptotic speed method uses the Big-O notation to describe the time computational complexity of a pedestrian model. Within the context of the computational speed of an algorithm, Big-O is defined as a function that describes how the computational effort of an algorithm increases as the input size of one particular input grows. More strictly, the Big-O method is defined as follow: 

Given a function $f(N)$, $O(f(N))$ represents the set of all $g(N)$ such that $|g(N)/f(N)|$ is bounded from above as $N \to \infty$ \cite{Sedgewick1996}. In other terms, it describes how the computational effort, described by $f(N)$, of an algorithm, grows with the size of the input ($N$) as an asymptotic approximation. For example, for an algorithm, the computational effort in relation to the size of its input ($N$) is captured by: $f(N) = 4N^2 + 6N + 10$. As $N$ (the input) goes to infinity, the term $N^2$ will dominate the other two terms. Hence, $f(N) = O(N^2)$.

This method is, in contrast to the empirical method, already well-defined in the literature. So, in this contribution, we will mainly focus on how one should apply the Big-O method to a pedestrian model in the context of our framework. The core of this method is to 1) find the loops and data structures that are related to the inputs and the size of all these inputs, and 2), determine the dominating structure and/or loops. To this end, an exact implementation of the algorithm is not necessary, pseudo-code with sufficient detail suffices as the model specification. 

To showcase how this method should be applied, we make use of a small example. The example is a basic implementation of a social force model described by the pseudocode in \autoref{alg:basic_sf_no_eq}. In the pseudocode of the example, we see one outer loop over all pedestrians and two inner loops (i.e. one loop over all pedestrians and one loop over all obstacles). Furthermore, there are no data structure operations other than getting a pedestrian or obstacle from a list which is a constant time operation. Using these insights, we can derive that the speed of this algorithm is described by $f(N,M) = aN^2 + bN + cMN + d$, where $N$ is the number of pedestrian, $M$ is the number of obstacles and $a-d$ are constants. From this equation, we can then derive that if the number of pedestrians goes to infinity ($N \to \infty$), the dominant term is $N^2$. Hence, the asymptotic speed of this algorithm is $O(N^2)$. That is, the growth rate of the speed is quadratic in relation to the number of pedestrians. Similarly, if the number of obstacles goes to infinity ($M \to \infty$), the term $cMN$ dominates the other terms and thus the speed is $O(M)$. That is, the growth rate is linear in relation to the number of obstacles.

\begin{algorithm}[H]
	\SetAlgoLined
	\ForEach{pedestrian}{
		Compute attraction force\\
		\ForEach{pedestrian $j$}{
			\If{$i \neq j$}{
				Compute pedestrian force\\
			}
		}
		\ForEach{obstacle}{
			Compute obstacle force
		}
		Compute the new velocity\\
		Compute the new position		
	}	
	\caption{Pseudocode describing a simple social force pedestrian model in continuous space.}
	\label{alg:basic_sf_no_eq}
\end{algorithm}

An important property of this method is that it is time-independent. So, for time-step-based models, the size of the time step does not impact the model’s growth rate within the definition of Big-O because it is a constant. For event-based models, such as the mesoscopic model by \cite{Tordeux2018}, the growth rate will depend on the frequency of the events. Yet, whether this frequency is defined as the number of events per second, per minute or any other unit of time is not relevant for the growth rate. As such, the method is also time-independent in this second case.   

It is important to note that an asymptotic approximation only holds for large input sizes. However, what this large input size is, is not defined by the method. This depends namely on the size of the constants and these are not determined within this method. Hence, this method only provides insight into a model's growth rate (i.e. is it linear, exponential etc. in relation to a certain input). This means that when two models have the same growth rate we cannot distinguish which model is faster even if one of the two would be much faster than the other if we would take the value of the constants into consideration. For example, if we have two models whose growth rates are linear in relation to the number of pedestrians ($N$) (i.e. $O(N)$) but whose actual speed is significantly different, for example $f(N) = 2N$ versus $f(N) = 1000N$, the first model is actually significantly faster (500 times) than the other. But, according to the asymptotic method, both models are similarly fast. 

It also means that if two models have different growth rates in relation to a certain input, we do not know for which ranges of the input size one is faster than the other and vice versa. Again, if we take two models but now with different growth rates in relation to the number of pedestrians, $O(N)$ versus $O(N^2)$ (i.e. linear versus quadratic), the first model is faster than the second, according to the asymptotic method. However, if we take their actual speed functions $f(N) = 1E^6N$ versus $f(N) = 2N^2$ we can see that for values of $N$ smaller than $\sqrt{1E^6/2} \approx 707 $, the model whose speed grows quadratically with the number of pedestrians will actually be faster.  

So, the asymptotic speed method only provides limited insights into the speed of pedestrian models. But if you are only interested in the question 'which model is likely faster when simulating very large scenarios' (i.e. the growth rate of the model), this is a very simple, and effective method to use, especially considering the other two, more tedious, methods in the framework.

\subsection{The atomic counting method}
\label{subsec:methods:atomic_counting}

The atomic counting method counts how many operations a single simulation step of a model takes. It makes use of the very detailed pseudocode of a model and the equations describing the model. This results in an equation describing a model's speed as a function of its inputs. To obtain the exact number of operations performed by a certain model implementation the Random Access Machine (RAM) model of computation is used \cite{Cook1973}.

\begin{figure}[htb]
	\centering
	\includegraphics{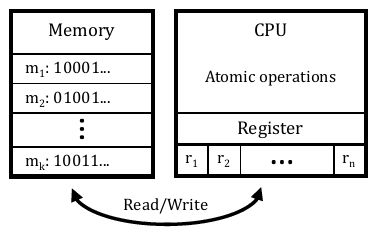}
	\caption{An overview of the word RAM model of computation}
	\label{fig:RAM_model}
\end{figure}

The RAM model represents a computing platform by means of three components \cite{Cook1973}, namely the CPU, the memory and the register (\autoref{fig:RAM_model}). The core of the RAM model is the CPU, which performs all computations and data manipulations. The memory and the register hold the data. The main difference between the latter two is that the CPU only has direct access to the data in the register (i.e. it can only manipulate or use data that is in the register). Each operation performed by the CPU is called an atomic operation.

In total there are four types of atomic operations:
\begin{enumerate}
	\item \emph{Register (re)initialization}: Set the register to a fixed value or to the value of another register (i.e. copying).
	\item \emph{Arithmetic operations}: Take the value of two registers and perform a basic arithmetic operation ($+$, $-$, $*$, $/$) using these values and store the result in the register.
	\item \emph{Comparison operations}: Take the value of two registers and compare them (e.g $=$, $>$, $\geq$) storing the result in the register or moving to the next operation based on the result (GOTO).
	\item \emph{Memory operations}: Read a value from memory into the register or write a value from the register to memory.
\end{enumerate}

The arithmetic and comparison operations are easy to count. It is harder to count memory operations because the number of memory operations depends on the number of registers, how often data has to be transferred from the register to and from the memory, and how often constants need to be initialized to ensure that there are enough registers available to complete the operation. The number of memory operations differs per computation platform and how each platform optimizes memory operations (also in combination with caches etc.). Therefore, we make the following assumptions:
\begin{enumerate}
	\item We assume the register size to be infinite (in line with the RAM model) during a single simulation step but not in between simulation steps. This means that every value has to be initialized or loaded from memory once during each simulation step and any value that should be saved should be written to the memory again.
	\item We presume that constant values that are used in every single simulation step (for example the value of the model time step) are persistent in the register and are loaded or initialized during the initialization of the model.
	\item We presume that a memory operation is a single value being read from memory, written to memory, or initialized in the register.
\end{enumerate}

Each operation also has its own cost. For most of the basic operations, this cost is roughly equal to 1 (unit cost) operation per CPU cycle. However, pedestrian models also commonly use more complex mathematical operations, such as logarithms, square roots, and trigonometric operations. These more complex mathematical operations are in essence small algorithms themselves that use basic (arithmetic) atomic operations to compute the result. Hence, these operations have a certain cost that is bigger than the unit cost of an atomic operation. 

The exact cost of more complex operations depends on multiple factors. The CPU, the compiler and the required numerical precision of the results, among other things. Therefore, it is impossible to provide exact numbers for the cost of the more complex operations that hold across all hardware and software. However, we can rank the operations from those who are generally the fastest to those who are generally the slowest. Furthermore, we can use the work from \cite{Fog2022}, which provides empirical measurements across a wide variety of CPU's for many different operations, to estimate the costs of more complex operations.

First, we set the cost of the basic arithmetic operations (addition, subtraction, multiplication, absolute) to 1 (unit cost). We do this as well for the memory operations and the comparison and logic operations. Division, modulo and rounding operations are generally somewhat slower than these basic operations. This is approximately by a factor of two, so we set the cost of these operations to 2. The square root operation is about 10 times slower than the basic operations (so the cost is 10) and computing logarithms and exponents is approximately two times slower than computing the square root (a cost of 20). The trigonometry operations are about 25 times slower than the basic operations and thus have  a cost of 25. Lastly, power operations are most expensive. About twice the cost of exponents so a cost of 40.

\autoref{tab:atomic_costs} presents an overview of the costs of all these common basic and more complex operations. These costs are, next to the assumption about the registers, another source of uncertainty in the atomic counting method. That is, they correctly represent the fact that some operations are computationally more expensive than others. However, by which factor precisely is uncertain.

\begin{table}[htb]
	\centering
	\caption{Overview of the cost of a collection of common operations}
	\label{tab:atomic_costs}
	\small
	\begin{tabular}{lRclR}
		\textbf{Operation(s)} & \textbf{Cost} && \textbf{Operation(s)} & \textbf{Cost}\\\hline
		Memory operations/Register initialization & 1 && Bit operations (e.g. right shift) & 1\\
		Comparisons and logic (e.g. if, $==$) & 1 && Basic math operations ($+$, $-$, $*$, $abs$) & 1\\			
		Division ($/$) & 2 && Modulo & 2\\
		Round, floor and ceil & 2 && Square root & 10\\
		Exponent & 20 && Logarithm & 20 \\ 
		Trigonometric operations (sin, atan, etc.)  & 25 && Power & 40\\
	\end{tabular}
\end{table}

To obtain the function describing the number of atomic operations in relation to the input of a specific algorithm, three steps need to be performed:

\noindent\emph{Step 1) Choose one specific implementation of the algorithm and write the accompanying pseudocode}: The method is independent of the computing platform, but it is not independent of the implementation of the model. Therefore, the first step is to write the detailed pseudo-code of a particular implementation of a pedestrian model and combine this with the model equations. This pseudocode and equations together form the model specification for this method.

\noindent\emph{Step 2) Per line count all the different atomic operations}: Per line of pseudocode, determine which atomic operations are performed, count them and relate them to the input size. For the memory operation (register initialization, memory load \& memory write) it is not always clear at which line the operation is performed. So, we advise to list all variables that are used in the algorithm and determine the number of memory related operation in that manner instead of trying to count them per line.

\noindent\emph{Step 3) Write the equation describing the model's speed based on the counts of the atomic operations}: Transform the count per line of pseudocode into a single equation describing the relation between the number of operations and the input size(s). Lastly, multiply the equation with the number of time steps per second ($1/\Delta t$). In the case of an event-based model this is the number of events per second.

To show how this method is applied to a model we use the example shown in \autoref{alg:simple_example_atomic}. These lines describe a part of the social force model, namely the loop over all pedestrians and the computation of the new position (\autoref{eq:example_atomic}) for each of those pedestrians. The pseudocode in \autoref{alg:simple_example_atomic} and \autoref{eq:example_atomic} together form the model specification in this example.

\begin{algorithm}[htb]
	\SetAlgoLined
	\ForEach{pedestrian}{
		Compute the position (according to \autoref{eq:example_atomic})\\ 
	}	
	\caption{Pseudocode describing a part of a social force model. Namely, the computation of the new position based on the newly computed velocity.}
	\label{alg:simple_example_atomic}
\end{algorithm}

\begin{equation}
	\vecut{p}{i}{t} = \vecut{p}{i}{t-1} + \vecut{v}{i}{t}\Delta t \label{eq:example_atomic} \\ 
\end{equation}
where $\vecut{p}{i}{t-1}$ is the position vector (x,y) for pedestrian $i$ at the previous time step ($t-1$), $\vecut{v}{i}{t}$ is the new velocity computed during this time step and $\Delta t$ is the time step. Furthermore, we assume that the velocity and the time step are already in the register and that the new position is written to memory because the next time step needs this to compute the next position. It is important to note that a loop is not an atomic operation but a set of atomic operations. These atomic operations includes, among other things, increasing the value of the loop counter every loop. In \autoref{subsec:complexities_loop} we present the atomic operations that make up a loop.

\autoref{tab:RAM_simple_example_counts_MEM} and \autoref{tab:RAM_simple_example_counts} present the cost of this algorithm for respectively the memory operation and the other (arithmetic and comparison) operations. If we combine the counts of these two tables, we get the following speed equation for this algorithm:
\begin{equation}
	f(N) = 8N + 2
\end{equation}

\begin{table}[htb]
	\centering
	\caption{The cost per memory operation for \autoref{alg:simple_example_atomic}. The memory operations, register initialization (init), read a value from memory (load) and write a value to memory (write). $N$ is the number of pedestrians.}
	\label{tab:RAM_simple_example_counts_MEM}
	\small
	\begin{tabular}{l l R }
		\textbf{Action} & \textbf{Type}  & \textbf{Cost}\\\hline
		Set the loop counter to 0 & Init &  1\\
		Load the loop size (\# peds.) into the register & Load & 1 \\
		Write the position to memory & Write & 2N \\
		\cline{3-3}
		\multicolumn{2}{l}{\textbf{Total costs}} &  \mathbf{2N + 2}\\
	\end{tabular}
\end{table}

\begin{table}[htb]
	\centering
	\caption{The cost per operation per line for \autoref{alg:simple_example_atomic}. $N$ is the number of pedestrians.}
	\label{tab:RAM_simple_example_counts}
	\small
	\begin{tabular}{l l R R R p{0.2\linewidth} }
		\textbf{Where} & \textbf{Operation}  & \textbf{Count} & \textbf{Cost per op.} & \textbf{Cost} & \textbf{Comments}\\\hline
		Line 1 & Addition & N & 1 & N & \\
		& Comparison & N & 1 & N & These operations are part of the loop (see \autoref{subsec:complexities_loop})\\
		Line 2 eq. (\ref{eq:example_atomic}) & Multiplications & 2N & 1 & 2N & $\vecut{v}{i}{t}*\Delta t$, one multiplication per part of the 2-D vector\\
		& Additions & 2N & 1 & 2N & $\vecut{p}{i}{t-1} + ...$, one addition per part of the 2-D vector\\
		
		\cline{5-5}
		\multicolumn{4}{l}{\textbf{Total costs}} &  \mathbf{6N} & \textbf{Operations}\\
	\end{tabular}
\end{table}

The main strength of the atomic counting method is that it provides a detailed insight into the speed of a pedestrian model. It can better distinguish between the speed of different models compared to the asymptotic method because it can distinguish between two models that have the same growth rate but very different complexities (i.e. $2N$ versus $1000N$, see the previous subsection). Additionally, this method provides insights into what part or parts of the given implementation of a model are most expensive (computationally) and thus could benefit from code optimization. Lastly, this method is independent of the computing platform. This can be especially useful when comparing models and model implementations cross-platform. For example, if models have a very similar theoretical computational speed but the empirical analysis shows a large difference, the atomic counting methods can identify which part of the slower model should be optimized for the given programming language.

The atomic counting method also has its weaknesses. Firstly, it requires a lot more effort to analyse an algorithm using the atomic counting method than the asymptotic method, especially when the algorithm and/or the used data structures are complex. In addition, the method requires some assumptions about the exact operation costs of non-atomic operations (e.g. logarithms, trigonometric operations etc.) and the number of memory operations in relation to the register size and use. Therefore, the value of the constants do contain some level of uncertainty. Hence, very small differences in model speeds, in relation to the input size, (i.e. a few operation per input size) cannot be considered to indicate a significant difference.

\subsection{How to choose a method or combination of methods}
\label{subsec:methods:choose}

The first step of the framework is to make the choice of which method or combination thereof you want to use for your analysis. To assist the users in making this choice we have summarized the main properties of each method in \autoref{tab:methods_overview}. This includes the core principle behind the method, the main output, its strengths, its limitations and an example of the type of application for which you would choose to use this method. As mentioned in \autoref{sec:framework} the best method is dependent on your particular goal and/or required analyses.

\begin{table}[htb]
	\centering
	\caption{An overview of the three methods including their main use cases, their strengths and their limitations}
	\label{tab:methods_overview}
	\small
	\begin{tabular}{p{0.1\linewidth} | p{0.28\linewidth} p{0.28\linewidth} p{0.28\linewidth}}
		& \textbf{Empirical} & \textbf{Asymptotic} & \textbf{Atomic counting}\\\hline
		\textbf{Core principle} & Measure the computation time of a model by running a single model (time) step for all test cases & Determine how the number of operations scales with the size of different inputs using the concept of the Big-O notation & Count the number of basic operations a computer must perform to execute a single (time) step of a model \\
		\textbf{Output} & A collection of run times for each model and set of test cases & The growth rate of the model in relation to the different input sizes in the form of Big-O notation & A detailed equation relating all inputs to the model's speed to the number of operations. \\
		\multirow[t]{2}{*}{\textbf{Strengths}} & Easy to use & Easy to use & Provides very detailed insights\\
		& Provides computation times & Independent of hardware and software & Independent of hardware and software\\
		\multirow[t]{3}{*}{\textbf{Limitations}} & Results depend very strongly on quality and level of optimization of implementation & Only provides valid results for very large scenarios (but does not tell how large a very large scenario is) & Is time-consuming and complicated to apply\\
		& Results depend strongly on used software and hardware & Low level of detail & Needs assumptions about the size and use of the registers and the cost of more complex operations\\
		& & No computation times & No computation times\\
		\textbf{Application} & Evaluating/comparing implemented model(s) in detail and and/or getting insight into computation times & Quick assessment of speed for (very) large scenarios
		& Evaluating/comparing model(s) in detail and independent of software and hardware.
	\end{tabular}
\end{table}

	\section{Test cases for determining the computational speed of pedestrian models}
\label{sec:test_cases}

The previous section shows that each of the three methods has its own strengths and weaknesses. In \autoref{sec:showcase} we will showcase the operationalization of the three methods in more detail. Yet, before doing so, we first need to establish a set of test cases, as the speed of a pedestrian model depends on the exact scenario that is simulated. For example, the speed of microscopic models strongly depends on the number of pedestrians in the scenario. On the other hand, the speed of cell-based macroscopic models strongly depends on the number of cells, which in turn is proportional to the size of the walkable space. Therefore, if one compares the computational speed of these two models only in relation to, for example, the size of the walkable space, the results will show that the microscopic model is much faster than the macroscopic model. And, while this is true in relation to the size of the walkable space, the opposite is true when one uses the number of pedestrians. Hence, this example shows that it is important to use multiple scenarios that jointly span the usage variation of the model to correctly capture the speed of a pedestrian model comprehensively; especially when comparing two or more models.

In this section, we present the part of the framework that captures the dependency of the speed of pedestrian models on the simulated scenarios in the form of test cases. Here, a test case represents a specific environment (e.g. walkable space shape and size) in combination with a specific traffic state (e.g. the number of pedestrians, their location, and their speed). A test case can be used to create the specific input for each of the pedestrian models of interest and compute/measure the speed for that specific test case. By creating sets of test cases and computing/measuring the model’s speed for each of these test cases you can obtain good insight into how the model is likely to perform in different scenarios.

We propose the following method to devise a set of test cases that will provide a robust overview of the impact of simulated scenarios on the speed of a pedestrian model of interest:

\subsection{Step A: Create the speed equation for each chosen method and model}
\label{subsec:test_cases:step_1}

To identify which elements of the environment and traffic state can impact a pedestrian model’s speed, we analyse the equation that describes the model’s speed. This speed equation is derived from the pseudocode in the case of the asymptotic and atomic counting methods. In the case of the empirical method, this equation is derived from the implemented code. 

For the asymptotic method, the speed equation is the equation that is derived as part of the method (see \autoref{subsec:methods:asymptotic} for the details). 

For the atomic counting method and empirical method, another procedure needs to be applied. The procedure is to identify any of the following operations from the model specification (i.e. code/pseudocode and equations see \autoref{sec:methods}):
\begin{itemize}
	\item Loops (For, while)
	\item Conditional statements (If)
	\item Data structure operations (that do not take constant time). For example, sorting, insertion into an array or searching a value in a data structure such as a KD-tree.
\end{itemize}  

To show how the speed equation is derived for all three methods we use an example. \AutorefAlg{alg:simple_macro_ctm} presents a small part of the pseudocode of a simple 1D unidirectional macroscopic cell transmission model (from now referred to as macroCTM). Namely, computing the demand (flux) from each cell to its neighbouring cell(s). Here $\rho_i$ is the density in cell $i$ and $\rho_{i;u}$ is the density of pedestrian class $u$ in cell $i$. \autoref{code:simple_macro_ctm} presents the same algorithm but then as implemented Python code for the empirical method. To keep the example simple, we do not provide the equations for line 5 of the pseudocode but assume it is an operation that is performed in some constant time.

\begin{algorithm}[H]
	\SetAlgoLined
	\ForEach{cell} {
		\If{$\rho_i > 0$}{
			\ForEach{pedestrian class} {
				\If{$\rho_{i;u} > 0$}{
					Compute the estimated demand for the neighbouring cell\\
				}
			}
		}
	}
	\caption{Pseudocode describing the first part of a simple implementation of a 1D unidirectional macroscopic cell transmission model}
	\label{alg:simple_macro_ctm}
\end{algorithm}

When we apply the asymptotic method to the pseudocode in \autoref{alg:simple_macro_ctm} we get the following equation:   
\begin{equation}
	\label{eq:example_asym_compl}
	f(L,K) = (bK + a)L + c
\end{equation} 
where $L$ is the number of cells, $K$ the number of pedestrian classes and $a-c$ are constants. Note that the conditional statements are not represented in \autoref{eq:example_asym_compl}. This is because the conditional statements do not impact the asymptotic growth rate but modify the magnitude of the constants and the magnitude of the constants is not relevant for the results of the asymptotic method.   

For the atomic counting method, the equation is similar but now includes the conditional statements. The equation is: 
\begin{equation}
	\label{eq:example_atomix_compl}
	f(L,K,\alpha,\beta) = (((d*\beta + c)K + b)\alpha + a)L + e
\end{equation} 
where $\alpha$ is the probability that the density in a cell is larger than 0 ($P(\rho_i > 0)$) and $\beta$ is the probability that the density in a cell for a certain pedestrian class is larger than 0 given that the density in the cell is larger than 0 ($P(\rho_{i;u} > 0 | \rho_i > 0)$). This equation is also the speed equation for the empirical method as the code in \autoref{code:simple_macro_ctm} shows that the (part of the) model is implemented with the same loops and conditional statements as the pseudocode which describes the model.

A speed equation is made up of three elements. The \emph{constants}, denoted by the small letter (e.g. $a$) define the constant cost of a certain set of operations. The \emph{asymptotic variables}, denoted by the capital letter (e.g. $L$), govern the asymptotic growth rate of the model’s speed. That is, all variables governing the size of the major loops and data structures. And lastly, the \emph{conditional modifiers}, denoted by the Greek lower-case letters (e.g. $\alpha$), that via some probability modify the number of times a certain set of operations is performed during a time step. We distinguish these three elements to make it easy to identify which variables need to be included in the test cases. The next few subsection will show how we use these elements to do create the test cases. 

\begin{code}
	\caption{A Python implementation of the first part of a simple implementation of a 1D unidirectional macroscopic cell transmission model (i.e. the implemented version of the pseudocode in \autoref{alg:simple_macro_ctm})}
	\label{code:simple_macro_ctm}
	\vspace{-5mm}
	\begin{minted}[breaklines=true,autogobble=true,mathescape,linenos,numbersep=5pt,frame=lines,framesep=2mm,python3=true,tabsize=4]{python}
    for i in range(cell_count):
        if density[i] > 0:
              for u in range(pedestrian_class_count):
                if density_per_class[i][u] > 0:
                # Compute the estimated demand for the neighbouring cell
	\end{minted}
\end{code}

\subsection{Step B: Identify the relevant model variables and their value ranges}
\label{subsec:test_cases:step_2}
The second step identifies which model variables impact the speed of a model. From the speed equation, it is easy to identify the model variables, these are namely the asymptotic variables and the conditional modifiers. The latter ones are only relevant when the atomic counting method or the empirical method is used. This step also involves determining the range of possible values for each variable. Again, this step is only relevant when the atomic counting method or the empirical method. In the case of the asymptotic method, the variables do not take a specific value but we determine the model's growth rate when these variables go to infinity.

For each of these model variables, a range must be provided by means of determining a lower and upper boundary. The lower boundary of the range is determined by the smallest non-empty scenario. For example, a single cell with a single pedestrian and thus a single pedestrian class. For the conditional modifiers, where the model variable is a probability, the upper bound is generally $1$. In the case of asymptotic variables, such as the number of cells, there is no clear upper bound as it could theoretically be infinite. As we show in the next step, also in those cases an upper bound has to be chosen when defining the test cases. \AutorefSubsec{subsec:test_cases:step_4} will show how we use this range to determine the values of each variables that make up the set of test cases.

If we take the macroCTM example and the speed equation derived in the previous part for the atomic counting method, we can derive four model variables. Two asymptotic variables in the form of the number of cells ($L$) and the number of pedestrian classes ($K$). And, two conditional modifiers. The first is the probability that a cell's density is larger than 0 ($\alpha$). The second is the probability that the density in a cell belonging to a certain pedestrian class is larger than 0, provided a cell's density is larger than 0 ($\beta$). \autoref{tab:value_ranges_example} provides an overview of these four variables and their value ranges. Below we explain in more detail how these value ranges are derived. 

The range of the number of cells is the set of positive integers (excluding 0) as any non-empty scenario will contain at least one cell, and in theory, can contain an infinite number of cells. For the number of pedestrian classes, this is also the case. Yet, in that case, the upper bound is limited by the number of pedestrians ($N$) in the simulation whereby $N = \rho*A$ with $\rho$ being the global density in the simulation and $A$ the total area of the walkable space. For both the conditional modifiers the upper boundary is $1$ which represents the case that all cells have a density larger than $0$ and in the case of $\beta$ all pedestrian classes are present in all cells. The lower boundary case for both is the case where only one cell has a density larger than 0 and this cell contains only a single pedestrian class. 

\begin{table}[htb]
	\centering
	\caption{Overview of the model variables from the example including their ranges and the underlying assumptions}
	\label{tab:value_ranges_example}
	\small
	\begin{tabular}{lp{0.28\linewidth}p{0.18\linewidth}p{0.11\linewidth}p{0.30\linewidth}}
		\multicolumn{2}{l}{\textbf{Variable}} & \textbf{Value range} & \textbf{Category} & \textbf{Assumptions}\\\hline
		$L =$ & The number of cells & $\mathbb{Z}^+$ & Asymptotic & The scenario is not empty and thus will contain at least a single cell\\
		$K =$ & The number of pedestrian classes & $\{k\in\mathbb{Z}^+ \;|\; k \leq N \}$ & Asymptotic & The number of pedestrian classes does not exceed the number of pedestrians in the simulation ($N$) \\
		$\alpha =$ & The probability that a cell has a density larger than 0 (i.e. $P(\rho_i > 0)$) & $[1/L,1]$ & Constant modifier & The scenario is not empty and hence will always contain at least 1 pedestrian and therefore at least one cell with a density larger than 0 (i.e. $\alpha*L \geq 1$)\\
		$\beta =$ & The probability that the density in a cell belonging to a certain pedestrian class is larger than 0, provided a cell's density is larger than 0 (i.e $P(\rho_{i;u} > 0 \;|\; \rho_i > 0)$) & $[1/K,1]$ & Constant modifier & Any cell with a density greater than 0 will at least contain a single pedestrian class (i.e. $\beta*K \geq 1$)\\
	\end{tabular}
\end{table}

\subsection{Step C: Generalize the model variables}
\label{subsec:test_cases:step_3}
Model variables describe elements of the context (i.e. space and state) of a pedestrian model in terms that can be specific to the model. Specific to the model means that the model variable is, in part, determined by model parameters. For example, for a cell-based model, the speed is, in part, determined by the number of cells variable. However, the number of cells is not a model-independent variable. It is a combination of the size and shape of the cell (a model parameter) and the size of the walkable space (model-independent).

To understand how the speed of a model relates to different simulation scenarios, it is important to separate the effect of the choice of the values of model parameters and the effect of the properties of the simulation scenarios. Therefore, the third step is to translate the model variable into these model-independent variables which we call state-space variables (because they describe the state and space of a pedestrian model independent of the model).

For example, for any cell-based model this means translating the number of cells $L$ into the size of the walkable space $A$. This is done using \autoref{eq:area_2_cell_count}. Here $a_\text{cell}$ is the surface area of the cell. Depending on the shape of the cell, how the cells are laid out in a grid and the geometry of the scenario, you might need more cells than you would expect strictly on the basis of dividing the surface area of the scenario by the surface are of a cell. The factors $\eta_\text{shape}$ compensates for this. It has a value of 1 in the case of a perfect fit (i.e. the walkable space is perfectly covered by the cell grid) and a value larger than 1 in all other cases.
\begin{equation}
    \label{eq:area_2_cell_count}
	L = \frac{A}{a_\text{cell}}*\eta_\text{shape}
\end{equation}

By defining the test cases in a common language for pedestrian models (i.e. the state-space variables), we can easily compare the results of a model with any other model. For example, the results of a cell-based model's speed are now presented in relation to the size of the scenario (i.e. the walkable space $A$) and it's shape ($\eta_\text{shape}$)  which can be easily compared to the speed of any other model in relation to the size and shape of the walkable space. This would not be the case if they were presented in relation to the number of cells.

\autoref{tab:model_to_scenario_variables} presents how the model variables of the macroCTM example are translated to their respective state-space variable or variables. Note that not all model variables translate to one single state-space variable. For example, the percentage of cells with a density larger than 0 is determined by three state-space variables, namely the spatial distribution of the pedestrians ($sd_p$), the number of pedestrians ($N$), and the size of the walkable space ($A$). The spatial distribution describes the difference between the global density in the scenario ($A/N$) and the average local density experienced by pedestrians ($k_\text{loc}$). If the spatial distribution is $0$ the average local density is equal to the global density which means the pedestrians are homogeneously distributed over the space. When the spatial distribution is $1$, pedestrians are as concentrated in the space as possible (i.e. they all experience the jam density ($k_\text{jam}$)). Please note, the number of pedestrian classes is a model variable that is not dependent on any model variable and thus also a state-space variable.

\begin{table}
	\centering
	\caption{Model variables to state-space variables (the number of pedestrian classes ($K$ is a model variable that is also already a state-space variable) )}
	\label{tab:model_to_scenario_variables}
	\small
	\begin{tabular}{p{0.10\linewidth}p{0.36\linewidth}cp{0.22\linewidth}}
		\textbf{Model variable} & \textbf{State-space variable(s)} & \textbf{Formula} & \\\hline
		$L$ &	Walkable space size ($A$) & $L = \frac{A}{a_\text{cell}}*\eta_\text{shape}$ & where $a_\text{cell}$ is the surface area of a cell and $\eta_\text{shape}$ the shape compensation factor\\
        		$\alpha$ & Spatial distribution of the pedestrians ($sd_p$) and the combination of the number of pedestrians and the size of the walkable space (i.e. the global density) & \makecell[t]{$\alpha=\frac{\left\lceil \frac{N}{a_\text{cell}*k_loc}\right\rceil*a_\text{cell}}{A}$ \\ \\ 
        			$k_\text{loc} = \frac{N}{A} + sd_p\left(k_\text{jam} - \frac{N}{A}\right)$} & where $k_\text{jam}$ is the jam density according to the specified fundamental diagram (a model parameter)\\
		$\beta$ & Spatial distribution of the pedestrian classes ($sd_k$) and the combination of the number of pedestrians and the size of the walkable space.	& $\beta = \frac{1}{K} + \frac{sd_k(K-1)}{K}$ &
	\end{tabular}
\end{table}	

\subsection{Step D: Create the test cases}
\label{subsec:test_cases:step_4}
Creating the test cases is a simple procedure for the asymptotic method. The set of test cases consists of the set of cases where each asymptotic state-space variable goes to infinity whilst the other asymptotic variables are still much smaller than infinity. The comprehensive set of test cases features the set of all possible unique combinations of asymptotic model variables going to infinity. If we take the example above, we identify that there are three test cases in total: 1) $K \to \infty$ and $A \ll \infty$, 2) $A \to \infty$ and $K \ll \infty$, and 3) $K \to \infty$ and $A \to \infty$.

For the other two methods (atomic and empirical), the procedure is different. This is necessary because there are in theory an infinite number of combinations of state-space variables values. This in turn leads to an infinite number of test cases. Therefore, we propose a procedure to create a base set of test cases that should be included. The base set of test cases is defined such that it provides insight into:
\begin{enumerate}[label=\alph*.]
	\item How each asymptotic state-space variable affects the model’s speed.
	\item How different combinations of values of the other variables impact the model's speed in relation to each asymptotic variable. 
\end{enumerate}
We prioritize the asymptotic state-space variables over the conditional modifiers because they define how the model’s speed grows as the simulated scenario grows (in space, number of pedestrians or population heterogeneity). In addition, they generally have a far larger range of values because they do not have a defined upper boundary. Therefore, it is important to know how the models’ speed relates to a wide range of values of these variables. To create this basic set of test cases we propose a procedure consisting of two steps: 1) Defining the test range for each asymptotic variable (step D.1). And 2), creating the actual test cases using these ranges (step D.2). In the remainder of these section we explain these steps and show how they are applied to the macroCTM example. Note that these step are only required when the atomic counting method or empirical method are used.

\subsubsection{Step D.1: Define the test range of the asymptotic variables}
This step involves choosing the upper and lower bound of the range and choosing the step size that discretizes the range. Asymptotic state-space variables generally have a range that, at least theoretically, can go to infinity and therefore they do not have a clearly defined upper boundary. Yet, in order to create a limited set of test cases, an upper bound has to be chosen. What this value should be is a matter of judgement. This can, for example, be guided by the largest possible values you expect for a particular application of the model. In general, we advise to start with a large value and lower this value if the upper part of the range does not add any information about either:
\begin{itemize}
	\item How the model’s speed relates to the value of the asymptotic state-space variable. This mainly relates to the shape of the curve that represent this relationship and whether this shape is stable.
	\item How the model’s speed in relation to the asymptotic state-space variable relates to the speed of another model for the same variable. This is mainly the question if all intersections between the curves, that represent the models' speeds in relation to the value of the asymptotic variable, fall within the range and thus can be observed.
\end{itemize}

The asymptotic variables usually have a clear lower boundary. However, at their smallest values, these asymptotic variables would describe very small scenarios. For example, a single pedestrian in a single cell. These cases are generally far smaller than any scenario to which a pedestrian model would be applied. Furthermore, these very small scenarios require very little computational effort regardless of, for example, their growth rate. So, we advise picking a lower boundary for the variables that fit a small but realistic scenario with values somewhat larger than their absolute minimum values. Again, what these values should be is a matter of judgement.

Each individual test case defines the input to a single simulation step of a model. Therefore, each test case is defined by only one value per state-space variable. Hence, we need a range of discrete values to define the test cases. Again, what a good discretization is, is a matter of judgement. It is a case of the smaller the better, however, for the empirical method, this might lead to a set of test cases that is too large to obtain their run times within a reasonable amount of time. 

\subsubsection{Step D.2: Create a set of test cases for each asymptotic variable}
To obtain insight into how an asymptotic variable affects a model's speed and how the values of other variables impact this relationship, we propose the following procedure to create a base set of test cases for each asymptotic variable:
\begin{enumerate}
	\item For each of the other variables select (at least) three values. We advice the use of at least three variables because this enables you to identify a trend of how this variables impacts the model's speed. For the conditional modifiers these three values would, by default, be the minimum value, the mean value and the maximum value of their range. For the other asymptotic variables, also at least three values must be chosen. These can be the minimum, maximum and mean value of the range defined in the previous step for these asymptotic variables. However, as these are not hard boundary values other values might be chosen for these variables if you see good reason to do so. 
	\item Take all the unique combinations of these variables and combine these with each value in the discretized range of the asymptotic variable. This forms a base set of test case for the asymptotic variable. That is, for each combination of variables, other than the asymptotic variable of interest, you get a subset of test cases where, if the model is applied to all these test cases, you get a relationship between the model speed and the asymptotic variable given that all other variables have a certain value.  
\end{enumerate}

Based on the speed results obtained by means of this base set of test cases, you can expand the set with additional test cases where, for example, you vary more than one asymptotic variable at a time, test different combinations of conditional modifier values or test a specific infrastructure with different states. Furthermore, when you compare models, we advise you to start with gaining insight into the speed of each individual model using these base test cases. But, if relevant, add test cases that, for example, show relevant differences in the complexities of the different models (if these are not already part of the base set).  

Using the macroCTM example, we show how this procedure to determine the test cases is applied. The first step is to define the value range for the two asymptotic variables, the size of the walkable space and the number of pedestrian classes. Here we choose a range between 50 and 250 with steps of 10 [$m^2$] for the walkable space area and a range of 1 to 50 [-] with a step size of 1 for the number of pedestrian classes. 

The next step is to choose, at least, three values per variable for both the asymptotic variables and the conditional modifiers. For the asymptotic variables, we choose the minimum and maximum value of the ranges defined in the previous step and a value in between. This leads to 50, 150 and 250 for the walkable space area and 1, 25 and 50 for the number of pedestrian classes.

For the two spatial distribution variables (pedestrians and pedestrian classes), we choose the minimum, mean and maximum values. These are 0, 0.5 and 1. For the global density, we take a slightly different approach. We choose values that relate to the different flow regimes, namely a density in a free flow ($k=0.5\: \mathrm{P/m^2}$) condition, the density at capacity ($k=1.25\: \mathrm{P/m^2}$) and the jam density ($k=5.4\: \mathrm{ P/m^2}$). Lastly, for the shape compensation factor ($\eta_\text{shape}$), we choose the minimum value of 1, a maximum value of 1.5 (i.e. you need 50\% more cells to cover the walkable space than in the case of a perfect fit due to the shape of the scenario and the cells) and 1.25 (the mean of the minimum and maximum value).

\autoref{tab:test_case_sets} presents an overview of the three chosen values per variable. By combining the variables into unique combinations per asymptotic variable, this results in $(A + K)*3^4 = (21 + 50)*81 = 5751$ individual test cases. This may seem like many simulations for the empirical method to run. But one must remember that every test case only requires a single time step to be run.

\begin{table}
	\centering
	\caption{The chosen three values for each of the six state-space variables.}
	\label{tab:test_case_sets}
	\small
	\begin{tabular}{RR|RRRR}
		\bm{K\: [-]} & \bm{A\: [m^2]} & \bm{\rho\: [ped/m^2]} & \bm{sd_p\: [-]} & \bm{sd_K\: [-]} & \bm{\eta_\text{shape}\: [-]}\\\hline
		1 &  50 & 0.5  & 0 & 0 & 1\\
		25 & 150 & 1.25 & 0.5 & 0.5 & 1.25\\
		50 & 250 & 5.4 & 1.0 & 1.0 & 1.5\\
	\end{tabular}
\end{table}

	\section{Showcasing the framework}
\label{sec:showcase}
To show how the framework works, we apply it to a naive social force model. \autoref{fig:framework} shows that the input to the framework are the model/models, and the goals and requirements of the analysis. Our goal is to showcase how to apply this framework to a pedestrian model. First, the naive social force model is introduced. Then all four steps are of the framework are applied to this model.

\subsection{The naive social force model}
\label{subsec:showcase:sf}

The model is a basic social force model \cite{Helbing1995} with a circular representation of the pedestrians. For simplicity, the fluctuation term is ignored and a static precomputed floor field is used for the route choice. Furthermore, each obstacle is a line or polygon made up of multiple lines whereby each individual line is seen as an obstacle. Lastly, a 0.1-second time step is used. \autoref{alg:basic_social_force} present the pseudocode of the model.

\begin{algorithm}[H]
	\SetAlgoLined
	\ForEach{pedestrian $i$}{
		Compute attraction force (eq. \ref{eq:bsf:dest})\\ 
		\ForEach{pedestrian $j$}{
			\If{$i \neq j$}{
				Compute pedestrian force (eq. \ref{eq:bsf:ped})\\
			}
		}
		\ForEach{obstacle}{
			Compute obstacle force (eq. \ref{eq:bsf:obs})\\
		}
		Compute the new velocity (eq. \ref{eq:bsf:vel})\\
		Compute the new position (eq. \ref{eq:bsf:pos})		
	}	
	\caption{Pseudocode describing a simple implementation of the social force model}
	\label{alg:basic_social_force}
\end{algorithm}

The algorithm of the model (\autoref{alg:basic_social_force}) is a loop over all pedestrians whereby the first step (line 2) in each loop is to compute the attraction force that attracts the pedestrian towards its goal using the following equation:
\begin{equation}
	\vecut{a}{i}{t} = \frac{\left(v_{i;0}\vecu{e}{i} - \vecut{v}{i}{t-1} \right)}{\tau_i} \label{eq:bsf:dest} \\
\end{equation}
where $\vecut{v}{i}{t}$ is the velocity of pedestrian $i$ at time $t$, $\vecut{a}{i}{t}$ the acceleration, $v_{i;0}$ the desired speed of the pedestrian (a parameter), $\vec{e}_\text{dir}$ the desired direction of the pedestrian obtained from the floor field and $\tau$ the relaxation parameter. For getting the vector of the desired direction from a static precomputed floor field we compute the x and y-coordinate based on the position of the pedestrian using:
\begin{equation}
	x = \lfloor p_x/l_{cell} \rfloor, \quad y = \lfloor p_y/l_{cell} \rfloor
	\label{eq:grid_pos}
\end{equation}
where $p_x$ is the x-coordinate of the position of the pedestrian and $l_{cell}$ is the cell size of the floor field. Using the x and y-coordinate the desired direction for that cell can be retrieved from memory.

The next step is computing the pedestrian interaction force working on pedestrian $i$ using an inner loop over all pedestrians (lines 3-5) and the following equations:
\begin{equation}
	\vecut{a}{i}{t} = \;\vecut{a}{i}{t} + \vecu{e}{j\to i}A_{p;i}e^{(r_i + r_j - d_{i;j})/B_{p;i}} \label{eq:bsf:ped} \\ 	
\end{equation}
\begin{subequations}
	\label{eq:p2p_vec}
	\begin{equation}
		\label{eq:p2p_vec:norm_vec}
		\vecu{e}{j\to i} = \frac{\vecu{p}{j \to i}}{d_{i;j}}  
	\end{equation}
	\begin{equation}
		\label{eq:p2p_vec:pos_diff}
		\vecu{p}{j \to i} = \vecu{p}{i} - \vecu{p}{j}
	\end{equation}
	\begin{equation}
		\label{eq:p2p_vec:dist}
		d_{i;j} = \sqrt{p_{j \to i;x}^2 + p_{j \to i;y}^2}
	\end{equation}
\end{subequations}
where $\vecu{e}{j\to i}$ is the vector pointing from pedestrian $j$ to pedestrian $i$, $A_p$ the pedestrian interaction force magnitude parameter, $r_i$ the pedestrian's radius, $d_{i \to j}$ the distance between the pedestrians and $B_{p;i}$ the pedestrian interaction force distance parameter.

The third step is to compute the obstacle force for each pedestrian using the inner loop over all obstacles and the following equation:
\begin{equation}
	\vecut{a}{i}{t} = \vecut{a}{i}{t} + \vecu{e}{o\to i}A_{o;i}e^{(r_i - d_{i \to o})/B_{o;i}}  \label{eq:bsf:obs} \\ 		
\end{equation}
where $\vecu{e}{o\to i}$ the vector pointing from obstacle $o$ to pedestrian $i$, $A_{o;i}$ the obstacle interaction force magnitude parameter, $d_{i \to o}$ the distance between the obstacle and the pedestrian and $B_{o;i}$ the obstacle interaction force distance parameter

The vector $\vecu{e}{o\to i}$ is the vector between the pedestrian's position and the closest point on the obstacle in relation to the pedestrian's position. The vector and the accompanying distance ($d_{i \to o}$) are computed using the same set of equations as \autoref{eq:p2p_vec}. As the obstacles are all lines the following algorithm and equations are used to compute the closest point on the obstacle:

\begin{algorithm}[H]
	\SetAlgoLined
	Compute the value of $\lambda$ (eq. \ref{eq:cp_lambda})\\
	\uIf{$\lambda < 0$}{
		$\lambda = 0$
	}
	\uElseIf{$\lambda > 1$}{
		$\lambda = 1$
	}
	Compute the closest point on the obstacle in relation to the pedestrian's position using $\lambda$ (eq. \ref{eq:cp_base:res})\\
	\caption{Pseudocode describing the algorithm to find the closest point on an line obstacle in relation to a pedestrian's position.}
	\label{alg:closest_point}
\end{algorithm}

\begin{subequations}
	\label{eq:closest_point}
	\setlength{\jot}{10pt}
	\begin{equation}
		\lambda = (\vecu{p}{i} - \vect{o}_{\text{org}}) \cdot \vect{o}_{\text{vec}}*ol_{sqr;inv} = ((p_x - l_{\text{org};x})o_{\text{vec};x} + (p_y - o_{\text{org};y})o_{\text{vec};y})*l_{sqr;inv} \label{eq:cp_lambda}
	\end{equation}
	\begin{equation}
		\vect{e}_{i \to i} = \vect{o}_{\text{org}} + \lambda*\vect{o}_{\text{vec}} \label{eq:cp_base:res}
	\end{equation}
\end{subequations}
We assume that the obstacle (i.e. the line) is described by its origin ($\vecu{o}{\text{org}}$), a vector ($\vecu{o}{\text{vec}}$) and the inverse of its squared length $l_{sqr;inv}$ (i.e. $1/||\vecu{o}{\text{vec}}||$).

The last step is to compute the new velocity and position of the pedestrian (lines 8 \& 9) using the Euler method and the following equations:
\begin{subequations}
    \label{eq:bsf:vel_pos}
	\setlength{\jot}{10pt}
	\begin{align}
		\vecut{v}{i}{t} =& \;\vecut{v}{i}{t-1} + \vecut{a}{i}{t}*\Delta t \label{eq:bsf:vel} \\ 
		\vecut{p}{i}{t} =& \;\vecut{p}{i}{t-1} + \vecut{v}{i}{t}*\Delta t \label{eq:bsf:pos}
	\end{align}
\end{subequations}
where $\Delta t$ the time step.

\subsection{Step 1: Selecting the method or methods}
Now that the model and goal are defined, the first step of the framework is to select the method or methods based on the goal. The goal is to showcase the framework. Therefore, all three methods are selected.

\subsection{Step 2: Providing the model specification}
Step 2 of the framework is to provide the model specification for each combination of a model and a method. In this example, it means that for each of the three methods, a model specification should be provided for the social force model. The model specification is the following for each method:
\begin{itemize}
    \item \textbf{Atomic counting method}: The detailed pseudocode (\autoref{alg:basic_social_force}) and the accompanying equations (equations \ref{eq:bsf:obs} to \ref{eq:bsf:vel_pos})
    \item \textbf{Asymptotic method}: The pseudocode (\autoref{alg:basic_social_force}). In this case you do not need the equations. Furthermore, line 4, a conditional statement, does not necessarily need to be included as conditional statements are not relevant for the asymptotic method.
    \item \textbf{Empirical method}: The implemented version of the pseudocode  (\autoref{alg:basic_social_force}) and equations  (\ref{eq:bsf:obs} to \ref{eq:bsf:vel_pos}). \autoref{code:basic_social_force} in \autoref{sec:source_code} shows the Python code that is used to run this model. 
\end{itemize}

\subsection{Step 3: Creating the test cases}
Creating the test cases requires four sub-steps. We walk through the four of steps one-by-one.

\subsubsection{Step 3.A: Creating the speed equations}
The first thing to do in this step is obtaining the speed equation of the model for each method. For the asymptotic method, we take the pseudocode in \autoref{alg:basic_social_force} and we identify that there are three loops and no data structure operations such as sort or search. From this we can derive the following speed equations:
\begin{equation}
	\label{eq:speed_asym}
	f(N,M) = (a + bN + cM)N + d
\end{equation}
where $N$ is the number of pedestrians, $M$ the number of obstacles and $a-d$ are constants. To deconstruct this, we provide the following overview of the cost per line or lines: 

\small
\noindent\begin{tabular}{lRRp{0.57\linewidth}}
	\textbf{Line(s)} & \textbf{Model variable(s)} & \textbf{Constant} & \textbf{Process(es)}\\
	1-2, 8-10 	& N & a & The cost that only depends on the outer loop such as the costs of computing the attraction force and the computation of the new velocity and position\\
	3-5 	& N^2	& b & The cost of computing the pedestrian interaction force for all $N*N$ combinations of pedestrians\\
	6-7 	& MN 	& c & The cost of computing the obstacle force for $M$ obstacles for all $N$ pedestrians\\
\end{tabular}

\normalsize
Finally, $d$ defines any costs that are only incurred once during a time step of the simulation (and thus not in one of the loops).

For the atomic counting method, we also use the pseudocode in \autoref{alg:basic_social_force} but now also include the conditional statements. The model contains three conditional statements. The if statement on line 4 and two if statements related to finding the closest point on an obstacle, line 2 and 4 of \autoref{alg:closest_point}. If we include these in the speed equation we get the following equation:
\begin{equation}
	\label{eq:speed_atomic}
	C(N,M,\alpha_{cp},\beta_{cp}) = (a + b(N - 1) + (c + d\acp + e(1 - \acp) + f\bcp)M)N + g
\end{equation}
This equation is very similar to the equation of the asymptotic method with two differences. Firstly, the term $bN$ has become $b(N-1)$ because of the if statement on line 4 whereby we know that this statement is false exactly once for every pedestrian. It must be noted that this statement does not add any new asymptotic variable or constant modifier to the equation due to the fact the probability of this statement being false is constant ($1/N$). The second difference ($cM$ versus $(c + d\acp + e(1 - \acp) + f\bcp)M$) does add two new variables to the equation. Namely, two conditional modifiers, $\acp$ and $\bcp$, which both relate to finding the closest point on an obstacle. $\acp$ is the probability that $\lambda < 0$ and $\bcp$ is the probability that $\lambda > 1$ 

The speed equation for the empirical method is equal to the equation for the atomic method in this case. That is, it has the same loops, data structures and conditional statements. They are equal because both methods use exactly the same implementation of the model.

\subsubsection{Step 3.B: Identifying the relevant variables and their value ranges}
The speed equations (\ref{eq:speed_asym} and \ref{eq:speed_atomic}) show that there are four relevant model variables in these equations. These are the two asymptotic variables, i.e. the number of pedestrians ($N$) and obstacles ($M$), and the two conditional modifiers, $\acp$ and $\bcp$. For each of these four variables, we need to define the range of values that they can take. \autoref{tab:value_ranges_social_force} shows all four variables with their accompanying value ranges and an explanation of these value ranges.

\begin{table}[htb]
	\centering
	\caption{Overview of the model variables of the social force model example including their ranges and the underlying assumptions}
	\label{tab:value_ranges_social_force}
	\small
	\begin{tabular}{lp{0.33\linewidth}p{0.11\linewidth}p{0.11\linewidth}p{0.30\linewidth}}
		\multicolumn{2}{l}{\textbf{Variable}} & \textbf{Value range} & \textbf{Category} & \textbf{Explanation}\\\hline
		$N$ & The number of pedestrians & $\mathbb{Z}^+$ & Asymptotic & The scenario is not empty and thus will contain at least a pedestrian and there is no strict upper boundary for the number of pedestrians.\\
		$M$ & The number of obstacles & $\mathbb{Z}^{0+}$ & Asymptotic & The scenario contains 0 or more obstacles and there is no strict upper boundary for the number of obstacles.\\
		$\acp$ & The probability that $\lambda$ is smaller than 0 $P(\lambda < 0)$ & $[0,1]$ & Conditional modifier & As it is a probability the range is naturally a value between 0 and 1.\\
		$\bcp$ & The probability that $\lambda$ is larger than 1 $P(\lambda > 1)$ & $[0,1-\acp]$ & Conditional modifier & The lower boundary of $\beta_{cp}$ is equal to 0 but the upper boundary is limited by the value of $\acp$.\\
	\end{tabular}
\end{table}

\subsubsection{Step 3.C: Generalizing the model variables}
All four of the model variables are model-independent. Therefore, for this naive social force model, we do not need to translate any of the variables.

\subsubsection{Step 3.D: Creating the test cases}
Now that we know all the relevant variables and their value ranges, we can create the test cases for all three methods. For the asymptotic method, this is simply the collection of all asymptotic variables and their unique combinations. So, for the social force model these are three test cases: \begin{enumerate}
	\item $N \to \infty$ and $M \ll \infty$: The relation between the number of pedestrians and the model's growth rate. 
	\item $M \to \infty$ and $N \ll \infty$: The relation between the number of obstacles and the model's growth rate.
	\item $N \to \infty$ and $M \to \infty$: The relation between both the variables and the model's growth rate.
\end{enumerate}

For the other two methods, we follow the steps detailed in \autoref{subsec:test_cases:step_3}. So, for each asymptotic variable, $N$ and $M$ in the case of the social force model, we first define a value range. For both the number of pedestrians and the number of obstacles, we choose a range of values between 5 and 200 with steps of 5. These ranges are relatively arbitrarily chosen, as we do not have a particular application in mind for which we want to test this model nor want to compare the model to another model. Hence, we choose the values such that they provide insight into the growth rates and the impact of each of the variables but also such that they are not so large and/or plentiful that they require a lot of computational effort when applying the empirical method. Furthermore, we choose the same range for both asymptotic variables so we can easily compare their impact.

The next step is to create the set of test cases for each of the asymptotic variables. First, we choose three values for each of the four variables. For the conditional modifiers these are the minimum, mean and maximum values if we follow our procedure described in \autoref{subsec:test_cases:step_3}. For the two asymptotic variables, we choose the values roughly the same way. We choose the minimum and maximum values of the ranges we defined in the previous step and a value more or less in between. \autoref{tab:test_case_values} presents an overview of these values.

\begin{table}[htb]
	\centering
	\caption{The three values per variable that are used to create the test cases}
	\label{tab:test_case_values}
	\begin{tabular}{r|RRRR}
		& \bm{N\; [-]} & \bm{M\; [-]} & \bm{\acp\; [-]} & \bm{\bcp\; [-]}\\\hline
		1 & 5 & 5 & 0.0 & 0.0\\
		2 & 100 & 100 & 0.5 & 0.5\\
		3 & 200 & 200 & 1.0 & 1.0\\
	\end{tabular}
\end{table}

According to our procedure, the set of test cases for each asymptotic variable is the combination of the range of that variable and the set of combinations of the values of all other variables. However, from the definition of the range of $\bcp$ we note that not all combinations of $\acp$ and $\bcp$ are valid as the upper bound of $\bcp$ is $1 - \acp$. So, if $\acp$ is 1, $\bcp$ is 0 by definition as both its lower and upper bound are 0.

Furthermore, we also note that the two conditional modifiers in this example only relate to a very small part of the operations performed by the model and thus a very small part of the cost. The $\acp$ and $\bcp$ variables both only impact to what degree the $d$, $e$ and $f$ constants (see \autoref{eq:speed_atomic}) contribute to the speed. Furthermore, from the pseudocode in \autoref{alg:closest_point} we can see that these constants are probably very small as they only represent the cost of a comparison operation or a register initialization operation. 

We additionally note that if $d$ and $e$ have the same value or have values that are close to each other, the impact of the value of $\acp$ on the model's speed will be very small. And we know that this is likely the case because the type of operations whose costs $d$ and $e$ describe are of very similar cost. In the case of the atomic counting method these are even exactly the same (namely unit cost). This also gives us reason to believe that in the empirical case these costs will be very similar. 

So, as the value of these conditional modifiers will likely impact the model's speed very little in comparison to the other variables, we treat them as a single variable when creating the test cases. That is, we do not use the 9 ($3x3$) combinations of values of these two variables but only three combinations. Namely, the case where they are both 0.5 (i.e. the average-case scenario), the case where $\acp$ is 1 and $\bcp$, by definition, 0 (i.e. the best-case scenario) and the case where $\acp$ is 0 and $\bcp$ is 1 (i.e the worst case scenario). This reduces the speed of the analysis by reducing the number of test cases per asymptotic variable. Because instead of the 27 ($3^3$) sets of relationships we get between an asymptotic variable and the model's speed we only get 9 ($3^2$). However, we still gain insight into what effect the conditional modifiers can have.

\autoref{tab:base_sets} presents the 9 sets of test cases per asymptotic variable. Together with the asymptotic variables which are both discretized into 40 values, this makes a total of 720 test cases ($(40 + 40)*9$).

\begin{table}[htb]
	\centering
	\caption{Overview of all the base sets of test cases per asymptotic variable}
	\label{tab:base_sets}
	\begin{tabular}{rRRRcrRRR}
		\mc{4}{c}{\textbf{\# pedestrians}} & & \mc{4}{c}{\textbf{\# obstacles}}\\
		\textbf{Set nr.} & M & \acp & \bcp && \textbf{Set nr.} & N & \acp & \bcp\\\hline
		1. & 5 & 0.5 & 0.5 && 1. & 5 & 0.5 & 0.5\\
		2. & 100 & 0.5 & 0.5 && 2. & 100 & 0.5 & 0.5\\
		3. & 200 & 0.5 & 0.5 && 3. & 200 & 0.5 & 0.5\\
		4. & 5 & 1 & 0 && 4. & 5 & 1 & 0\\
		5. & 100 & 1 & 0 && 5. & 100 & 1 & 0\\
		6. & 200 & 1 & 0 && 6. & 200 & 1 & 0\\
		7. & 5 & 0 & 1 && 7. & 5 & 0 & 1\\
		8. & 100 & 0 & 1 && 8. & 100 & 0 & 1\\
		9. & 200 & 0 & 1 && 9. & 200 & 0 & 1\\
	\end{tabular}
\end{table}

\subsection{Step 4: Applying all methods and test cases to the model}

In Step 4, we apply all three methods to all their respective test cases. The results of this exercise are presented per method.

\subsubsection{Asymptotic method}
For each of the three test cases of the asymptotic method, we derive the dominating term or terms and with that the asymptotic speed. \autoref{tab:asymptotic_method_res} presents the results for each of the test cases. From these results, we observe that the model's computational effort grows quadratically with the number of pedestrians ($N$) and linearly with the number of obstacles ($M$). In the case that the value of both variables goes to infinity, the model computational effort is still only dominated by the number of pedestrians. That is, if the number of pedestrians goes to infinity, the number of obstacles does not impact the growth rate even if its value also goes to infinity. This is due to the additive nature of the two inner loops in the outer loop ($(bN + cM)N$). In other words, if both values go to infinity, $N \simeq M$ so $(bN + cM)N$ would be similar to $(bN + cN)N$ which is equal to $(b+c)N^2$.

Therefore, we can conclude that the speed of this implementation of the naive social force model is especially sensitive to the number of pedestrians and to a lesser degree sensitive to the number of obstacles in the case of large scenarios with many pedestrians. However, in small but complex scenarios (i.e. a few pedestrians but many obstacles) either one could be dominant depending on the size of the constants, Yet, these constants are not provided by this method. Nor does the method tell us what constitutes a large scenario and what constitutes a small scenario.

\begin{table}[htb]
	\centering
	\caption{The application of the asymptotic method to the three test cases}
	\label{tab:asymptotic_method_res}
	\begin{tabular}{RRR}
		\mc{1}{l}{\textbf{Test case}} & \mc{1}{l}{\textbf{Dominant term}} & \mc{1}{l}{\textbf{Asymptotic speed}}\\
		N \to \infty \text{ and } M \ll \infty & bN^2 & O(N^2)\\
		M \to \infty \text{ and } N \ll \infty & cM & O(M)\\
		N \to \infty \text{ and } M \to \infty &  bN^2 & O(N^2)\\
	\end{tabular}
\end{table}

\subsubsection{Atomic counting method}

From the pseudocode in \autoref{alg:basic_social_force} we derived the following speed equation by applying the atomic counting method:
\arraycolsep=1.4pt
\begin{equation}
	\small
	\label{eq:ssf_atomic}
	\begin{array}{rcrclrcrcrcrcrcr}
		\text{FLOPS}  & = & 51N^2 & + & ( & 61 & - & \acp &   &     )MN & - & 70N &   &   & + & 48\\
		\text{MEMOPS} & = &    	  &   & ( &    &   & \acp & + & \bcp)MN & + & 19N & + & 5M & + & 3\\
		\cline{3-16}
		\text{Total}  & = & 51N^2 & + & ( & 61 &   &      & + & \bcp)MN & - & 51N & + & 5M & + & 51\\
	\end{array}
\end{equation}

For a detailed line-by-line and equation-by-equation derivation of this equation, we refer to \autoref{app:atomic_sf_details}. The equation shows a few things. Firstly, the conditional modifier $\acp$ is cancelled out and thus its value does not impact the model’s speed. Furthermore, the impact of the other conditional modifier, $\bcp$, is also very small as it only determines the likelihood that a single operation is performed and this operation only has a cost of 1. So, we can conclude from the equation that the exact values of the conditional modifiers are not very relevant to the model's speed. 

We also observe that the number the computations (FLOPS) are the main source of the model's speed and that the memory operations (MEMOPS) contribute less. We can also easily derive the model's growth rates in relation to the two asymptotic variables. The equations show that the model's speed will grow quadratically with the number of pedestrians ($N$) and linearly with the number of obstacles ($M$). We also note that the sizes of all constants are within the same order of magnitude. That is, no part of the equation (and with that no part of the model) will dominate the other parts in the case of small scenarios (small $N$ and $M$).

\begin{figure}[htb]
	\centering
	\includegraphics{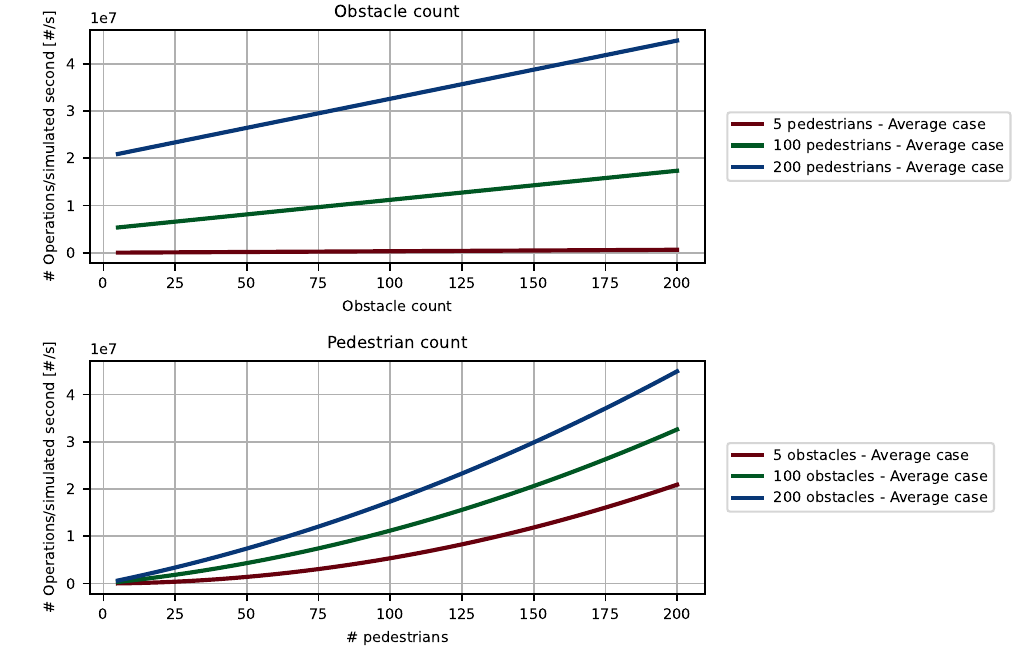}
	\caption{The relationship between the social force model's speed and the number of pedestrian and obstacles obtained using the atomic counting method}
	\label{fig:results_atomic}
\end{figure}

We also visualized the equation by plotting different sets of test cases. \autoref{fig:results_atomic} presents the results. The curves are plugging in the relevant values of all four variables into \autoref{eq:ssf_atomic}. Note that we only plotted the average case scenarios (sets 1-3 from \autoref{tab:atomic_costs}) as we already concluded that the values of the conditional modifiers have no or very little impact. In addition, note that the results are presented as the number of operations per simulated second. That means the number of operations are multipied by 1 over the time step of the model (i.e. $1/\Delta t$). The graphs especially show the effect the value of the asymptotic variables have on each other. 

\subsubsection{Empirical method}
\begin{figure}[htb]
	\centering
	\includegraphics{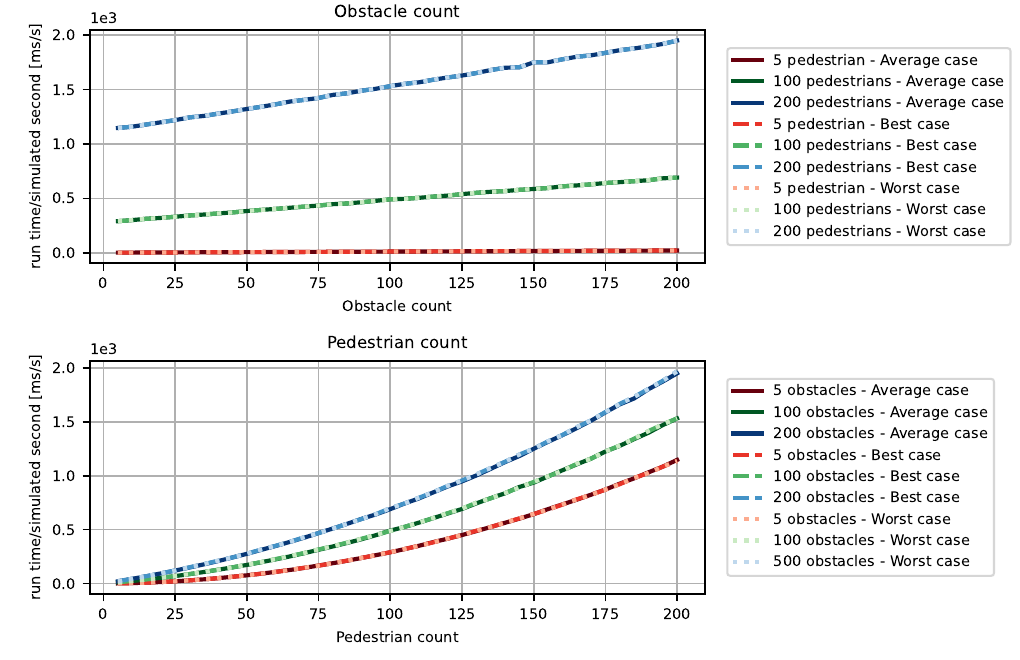}
	\caption{The relationship between the social force model's speed and the number of pedestrian and obstacles obtained using the empirical method}
	\label{fig:results_empirical}
\end{figure}

For the empirical method, we ran all 720 experiments (720 test cases x 1 model). The mean run time of all experiments converged within 60 replications (with a minimum of 30 replications). \autoref{fig:results_empirical} presents the result per asymptotic variable.

The graphs show very similar results to the results in \autoref{fig:results_atomic} from the atomic method. They show the quadratic growth in the model's speed in relation to the number of pedestrians and a linear growth in relation to the number of obstacles. Furthermore, the results also show that the values of the two conditional modifiers do not significantly impact the model's speed.	

	\section{Conclusions and discussions}
	\label{sec:conclusions}
	In this paper, we presented a framework to assess the computational speed (formally the time computational complexity) of pedestrian models in a systematic manner. The core feature of this framework is that it leverages three methods jointly to cover the various goals a scientist or engineer could have when assessing a model's speed. These methods are the Empirical method, the Asymptotic method and the Atomic counting method. In addition, this new framework provides a procedure to systematically create a set of test cases that cover all relevant scenarios that might impact the model's speed.
	
    \AutorefSec{sec:showcase} has shown that the selected methods each have their own strengths and limitations. Where the empirical method is more useful in cases where model implementations are compared, the asymptotic method is more useful to compare model types quickly. The Atomic counting method provides in-depth knowledge regarding the relation between specific algorithmic steps and the computational speed of a comprehensive model. This paper has developed a guideline to determine which method can best support researchers and engineers to derive insights regarding computational speed at various stages of model development.
	
	The newly designed procedure for creating the set of test cases provides a systematic way to derive and include all variables that impact a model's speed. This ensures that the set of test cases, that is used to test a model's speed, covers the full spectrum of scenario elements that determine a model's speed. This prevents that incomplete and/or biased insights into a model's speed are obtained, which is especially relevant when models are compared. In this case, an incomplete set of test cases and the subsequent incomplete insights into the speed of the models can lead to wrong conclusions about which model is faster in which case. Like the methods, the procedure is also (mostly) not specific to pedestrian models.   

    We expect that the application of this framework is not limited to pedestrian models. Our adaptations of the atomic counting method and the empirical method, basing it on a single time step (or event-based equivalent) and converting the results to the computational effort per simulated second, make that this framework (featuring these three methods) are applicable to any time-based simulation model. Pedestrian models are just one category of this type of models. Traffic models, and agent-based models featuring logistics are just two other examples that could benefit from this new framework.
	
	\subsubsection{Limitations}
	The framework has a few limitations which we like to discuss in more detail. Firstly, the framework limits itself to the time computational complexity of pedestrian models, thereby ignoring the space computational complexity (i.e. the memory usage) of these models. Though we expect that the amount of computer memory a pedestrian model uses will generally not be a limiting factor, it still can be the case that it is for very large scenarios or for hardware with limited memory capacity. It is also the case that endeavours aimed at increasing the speed of an algorithm can result in an increase the memory consumption of the model. For example, algorithms that store intermediate results instead of recomputing them. Therefore, future research into this part of pedestrian models' computational speed can be worthwhile to obtain more insight into this trade-off between computational speed and memory usage. Lastly, the ratio between the number of memory operations and floating-point operations can also impact a model's performance \cite{Ofenbeck2014}.  
	
	Secondly, with this speed framework, we focus solely on the walking behaviour part of a pedestrian model. To obtain insight into the speed of a fully functional pedestrian model, the speed of all other elements, such as the route choice behaviour model or the source and sink algorithms, also needs to be analysed to get a complete picture of the speed. Though this framework does not focus on these elements, it can most likely, with some small adaptions, be applied to these other elements. However, future research should study this to prove that this is indeed the case and to show what adaptations are necessary. 

    Thirdly, each of the three methods has it's own limitations. The Asymptotic method only provides growth rates and is only valid for large input sizes. The results of the Empirical method depend strongly on the used computing platform. So, to compare models or model implementations in detail, the same exact computing platform must be used. Results obtained using different platform can be compared, but only to a limited degree. Lastly, in the Atomic counting method there is some uncertainty about the exact cost of more complex operations and the size of the register. Therefore, small differences might not be significant due to these uncertainties. In short, no method is perfect and in all cases you should explicitly take into account its limitations when interpreting the results.

    Lastly, this paper introduces our new framework and therefore focuses on a detailed description of the methods and the test case creation procedure. We also provide one example of how this new framework can be applied to a pedestrian model. Future research can take a more detailed look at how the framework can be applied to different models in different contexts. Or show how it can be applied to compare different models to each other. Lastly, future research can use our new test case creation procedure to systematically study which scenario elements impact the speed of which type of model.
	
	\printbibliography
	
	\appendix
	
    \section{The atomic speed derivation of the social force model}
\label{app:atomic_sf_details}

\begin{table}[htb]
	\caption{A line by line and equation by equation count of the number of arithmetic and logical operations performed by the social force model presented in \autoref{subsec:showcase:sf}}
	\label{tab:flops}
	\begin{opCountTabuMul}
		1	& Loop  		& 1 	& 2		& N      	& 2N 		& See \autoref{subsec:complexities_loop} \\
		2	& Sub   		& 2     & 1		& N     	& 2N    	& \refEq{eq:bsf:dest}\\
			& Mul  			& 2     & 1		& N     	& 2N    	& \refEq{eq:bsf:dest}\\
			& Div   		& 4     & 2		& N     	& 8N    	& \refEq{eq:bsf:dest} (2x) + \refEq{eq:grid_pos} (2x)\\
			& Floor   		& 2     & 3		& N     	& 6N    	& \refEq{eq:grid_pos}\\
		3	& Loop  		& 1 	& 2		& N^2      	& 2N^2 		& Squared due to it being an inner loop.\\
		4	& Comp  		& 1     & 1		& N^2   	& N^2   	&  \\
		5	& Add   		& 4     & 1		& (N-1)^2 	& 4(N-1)^2 	& \refEq{eq:bsf:ped} (3x) + \refEq{eq:p2p_vec:dist} (1x)\\
			& Sqrt   		& 1     & 10	& (N-1)^2 	& 10(N-1)^2 & \refEq{eq:p2p_vec:dist}\\
			& Sub   		& 3     & 1		& (N-1)^2 	& 3(N-1)^2 	& \refEq{eq:bsf:ped} (1x) + \refEq{eq:p2p_vec:pos_diff} (2x)\\
			& Mul   		& 5     & 1		& (N-1)^2 	& 5(N-1)^2 	& \refEq{eq:bsf:ped} (3x) +  \refEq{eq:p2p_vec:dist} (2x)\\
			& Div   		& 3     & 2		& (N-1)^2 	& 6(N-1)^2 	& \refEq{eq:bsf:ped} (1x) + \refEq{eq:p2p_vec:norm_vec} (2x)\\
			& Exp   		& 1     & 20	& (N-1)^2 	& 20(N-1)^2 & \refEq{eq:bsf:ped}\\
		6	& Loop 	 		& 1 	& 2		& MN      	& 2MN  		& Multiplied by $N$ due to it being an inner loop \\
		7	& Add   		& 6     & 1		& MN    	& 6MN   	& \refEq{eq:bsf:obs} (2x) + \refEq{eq:cp_lambda} (1x) + \refEq{eq:cp_base:res} (2x) + \refEq{eq:p2p_vec:dist} (1x)\\
			& Sqrt    		& 1	    & 10	& MN   		& 10MN 		& \refEq{eq:p2p_vec:dist}\\
			& Sub   		& 5     & 1		& MN    	& 5MN    	& \refEq{eq:bsf:obs} (1x) + \refEq{eq:cp_lambda} (2x) + \refEq{eq:p2p_vec:pos_diff} (2x) \\
			& Mul  		 	& 10    & 1		& MN    	& 10MN   	& \refEq{eq:bsf:obs} (3x) + \refEq{eq:cp_lambda} (3x) + \refEq{eq:cp_base:res} (2x) + \refEq{eq:p2p_vec:dist} (2x)\\
			& Div  		 	& 3     & 2		& MN    	& 6MN   	& \refEq{eq:bsf:obs} (1x) + \refEq{eq:p2p_vec:norm_vec} (2x)\\
			& Exp   		& 1     & 20	& MN    	& 20MN   	& \refEq{eq:bsf:obs}\\
		2 (Alg. \ref{alg:closest_point}) & Comp & 1 & 1 & MN & 1MN & \\
		4 (Alg. \ref{alg:closest_point}) & Comp & (1 - \acp) & 1 & MN & (1 - \acp)MN & $(1 - \acp)$ is the probability that $\lambda \leq 0$\\
		8	& Add  		 	& 2     & 1		& N     	& 2N    	& \refEqP{eq:bsf:vel}\\
			& Mul   		& 2     & 1 	& N     	& 2N    	& \refEqP{eq:bsf:vel}\\		
		9	& Add  		 	& 2     & 1		& N     	& 2N    	& \refEqP{eq:bsf:pos}\\
			& Mul   		& 2     & 1		& N     	& 2N    	& \refEqP{eq:bsf:pos}\\
		\mc{6}{R}{\textbf{Totals = }\quad \bm{3N^2 + 48(N-1)^2 + (60 - \acp)MN + 26N}} & \textbf{operations}
	\end{opCountTabuMul}
\end{table}

\begin{table}[htb]
	\caption{The number of memory operations performed by the social force model presented in \autoref{subsec:showcase:sf}}
	\label{tab:memops}
	\begin{memopCountTabu}
		Load pedestrian state & Load & 4 & N & 4N & Load the position (2 values) and velocity (2 values) of all pedestrians into the register\\
		Load pedestrian parameters & Load & 7 & N & 7N & Load all parameters for all pedestrians into the register ($\tau_i, v_{0;i}, A_{p;i}, ri, B_{p;i}, A_{o;i}, B_{o;i}$)\\
		Load desired direction from floor field & Load & 2 & N & 2 N & Load the desired direction from the floor field ($\add$ is the probability that the direction has not yet been loaded for another pedestrian in the same floor field cell)\\
		Load obstacles & Load & 5 & M & 5M & Load all obstacles into the register. An obstacle is defined by a list with 5 entries.\\
		Load the loop lengths & Load & 2 & & 2& Load the length of the three loops (the number of pedestrian for two of the loops and the number of obstacles)\\
		Loop counter init outer pedestrian loop & Init & 1 & & 1 & Set the loop counter to 0\\
		Loop counter init inner pedestrian loop and the obstacle loop & Init & 2 & N & 2N & Set the loop counter to 0 once every time the loop is performed\\
		Set $\lambda$ to 0 or 1 & Init & (\acp + \bcp) & MN & (\acp + \bcp)MN & Where $\acp$ is the probability that $\lambda < 0$ and $\bcp$ is the probability that $\lambda > 1$ \\
		Write pedestrian state & Write & 4 & N & 4N & Write the position and velocity of all pedestrians from the register back into the memory\\
		\mc{5}{R}{\textbf{Totals = }\quad \bm{(\acp + \bcp)MN + 19N + 5M + 3}} & \textbf{operations}
	\end{memopCountTabu}
\end{table}

\section{The speed of basic operations}
\label{app:basic_ops}

In this appendix we present some of the basic operations used in many pedestrian models or in one of the examples. We also analyse and present their speed according to the atomic counting method. 

\subsection{A loop}
\label{subsec:complexities_loop}

A loop is not a single operation but actually a set of operations. \AutorefAlg{alg:loop} shows how the pseudocode of a loop with all the operation that are performed during the loop.  

\begin{algorithm}[H]
	\SetAlgoLined
	Initialize counter i = 0\\
	Load n from memory into the register\\
	\Repeat{i > n}{
		Perform some operations in the loop\\
		i =  i + 1
	}	
	\caption{Pseudocode describing a loop}
	\label{alg:loop}
\end{algorithm}

The number of operations is as follow:\\
\begin{opCountTabu}
	\textbf{Line} & \textbf{Count} & \textbf{Operation} & \textbf{Costs} & \textbf{Details}\\
	\hline
	\multicolumn{5}{l}{\textbf{FLOPS}}\\
	\hline
	5 & n & Additions & n &\\
	6 & n & Comparisons & n &\\
	\cline{4-4}
	\multicolumn{3}{r}{\textbf{Totals}} & \bm{2n} & \textbf{operations}\\ 	
	\mc{5}{l}{where $n$ is the size of the loop.}\\
	\multicolumn{5}{l}{\textbf{MEMOPS}}\\
	\hline
	1 & 1 & Register init & 1 &\\
	2 & 1 & Memory load & 1 &\\
	\cline{4-4}
	\multicolumn{3}{r}{\textbf{Totals}} & \bm{2} & \textbf{operations}\\ 	
\end{opCountTabu}\\

\section{Source code social force model}
\label{sec:source_code}

\begin{code}
	\caption{The python code for the basic social force model. For the code of the get\_pref\_dir and closest\_point\_on\_line functions see below}
	\label{code:basic_social_force}
	\begin{minted}[breaklines=true,autogobble=true,mathescape,linenos,numbersep=5pt,frame=lines,framesep=2mm,python3=true,tabsize=4,fontsize=\small]{python}
		for ped in self.pedestrians:
			pref_dir_x, pref_dir_y = self.get_pref_dir(ped.pos_x, ped.pos_y)
			acc_x = ped.relaxation_time_inv*(ped.pref_speed*pref_dir_x - ped.vel_x)
			acc_y = ped.relaxation_time_inv*(ped.pref_speed*pref_dir_y - ped.vel_y)
			
			for other_ped in self.pedestrians:
				if ped == other_ped:
					continue
				
				x_diff = ped.pos_x - other_ped.pos_x
				y_diff = ped.pos_y - other_ped.pos_y
				dist = sqrt(x_diff*x_diff + y_diff*y_diff)
				force_mag = ped.ped_force_mag*exp((ped.radius + other_ped.radius - dist)/ped.ped_force_sig)
				acc_x += x_diff/dist*force_mag
				acc_y += y_diff/dist*force_mag
			
			for line in self.obstacles:
				closest_x, closest_y = closest_point_on_line(line, ped.pos_x, ped.pos_y)
				x_diff = ped.pos_x - closest_x
				y_diff = ped.pos_y - closest_y
				dist = sqrt(x_diff*x_diff + y_diff*y_diff)
				force_mag = ped.obs_force_mag*exp((ped.radius - dist)/ped.obs_force_sig)
				acc_x += x_diff/dist*force_mag
				acc_y += y_diff/dist*force_mag
			
			ped.vel_x += acc_x*self.DELTA_T
			ped.vel_y += acc_y*self.DELTA_T # type: ignore
			
			ped.pos_x += ped.vel_x*self.DELTA_T
			ped.pos_y += ped.vel_y*self.DELTA_T	
	\end{minted}
\end{code}

\begin{code}
	\caption{The python code for getting the preferred direction for a pedestrian at (pos\_x, pos\_y) from a precomputed static floor field}
	\label{code:des_dir}
	\vspace{-5mm}
	\begin{minted}[breaklines=true,autogobble=true,mathescape,linenos,numbersep=5pt,frame=lines,framesep=2mm,python3=true,tabsize=4]{python}
		row_ind = int(floor(pos_y/ROUTE_CELL_SIZE))
		col_ind = int(floor(pos_x/ROUTE_CELL_SIZE))	
		pref_dir_x, pref_dir_y = self.pref_direction[row_ind, col_ind,0], self.pref_direction[row_ind, col_ind,1]
	\end{minted}
\end{code}

\begin{code}
	\caption{The python code for getting the preferred direction for a pedestrian at (pos\_x, pos\_y) from a precomputed static floor field}
	\label{code:closest_point}
	\vspace{-5mm}
	\begin{minted}[breaklines=true,autogobble=true,mathescape,linenos,numbersep=5pt,frame=lines,framesep=2mm,python3=true,tabsize=4]{python}
		# line = [[org_x, org_y], [vec_x, vec_y], length_sqr_inv]
		rel_line_pos = ((pos_x - line[0][0])*line[1][0] + (pos_y - line[0][1])*line[1][1])*line[2]
		if rel_line_pos < 0:
			rel_line_pos = 0
		elif rel_line_pos > 1:
			rel_line_pos = 1
		
		x = line[0][0] + rel_line_pos*line[1][0]
		y = line[0][1] + rel_line_pos*line[1][1]
	\end{minted}
\end{code}

\end{document}